\documentclass[epjST]{svjour}
\usepackage{graphics}
\usepackage{amsmath,amssymb,graphicx,color}

\newcommand{\mean}[1]{\left\langle #1 \right\rangle}

\def\E{{\mathcal E} }
\def\llangle{\left\langle}
\def\rrangle{\right\rangle}

\begin{document}
\title{Collectivity in large and small systems formed in ultrarelativistic collisions}
\author{Rajeev S. Bhalerao\inst{1}
\fnmsep\thanks{\email{rajeev.bhalerao@iiserpune.ac.in}} 
}
\institute{$^1$Department of Physics, Indian Institute of Science Education and Research (IISER), Homi Bhabha Road, Pune 411008, India 
}
\abstract{ Collective flow of the final-state hadrons observed in
  ultrarelativistic heavy-ion collisions or even in smaller systems
  formed in high-multiplicity pp and p/d/$^3$He-nucleus collisions is
  one of the most important diagnostic tools to probe the initial
  state of the system and to shed light on the properties of the
  short-lived, strongly-interacting many-body state formed in these
  collisions. Limited, in the initial years, to the study of mainly
  the directed and elliptic flows -- the first two Fourier harmonics
  of the single-particle azimuthal distribution -- this field has
  evolved in recent years into a much richer area of activity. This
  includes not only higher Fourier harmonics and multiparticle
  cumulants, but also a variety of other related observables, such as
  the ridge seen in two-particle correlations, flow decorrelation,
  symmetric cumulants and event-plane correlators which measure
  correlations between the magnitudes or phases of the complex flows
  in different harmonics, coefficients that measure the nonlinear
  hydrodynamic response, statistical properties, such as the
  non-Gaussianity of the flow fluctuations, etc. We present a Tutorial
  Review of the modern flow picture and the various aspects of the
  collectivity -- an emergent phenomenon in quantum chromodynamics.  }
%
\maketitle
%
\newpage

\vspace*{5cm}

``The aim of theory really is, to a great extent, that of systematically
organizing past experience in such a way that the next generation, our
students and their students and so on, will be able to absorb the
essential aspects in as painless a way as possible, and this is the
only way in which you can go on cumulatively building up any kind of
scientific activity without eventually coming to a dead end.''

\bigskip

--- Michael Atiyah, Fields Medalist, in {\it How research is carried out}

\newpage

\noindent Contents

\noindent1. Introduction\\
2. Modern Flow Picture\\
\hspace*{0.3cm} 2.1 Characterization of the initial state\\
\hspace*{0.3cm} 2.2 Characterization of the final state\\
\hspace*{0.3cm} 2.3 Flow measurements\\
\hspace*{0.3cm} 2.4 Probability density function (PDF)\\
\hspace*{0.3cm} 2.5 Ridge\\
\hspace*{0.3cm} 2.6 Flow decorrelation and factorization breaking\\
3. Observables with Mixed Harmonics\\
\hspace*{0.3cm} 3.1 Symmetric cumulants\\
\hspace*{0.3cm} 3.2 Event-plane correlators\\
\hspace*{0.3cm} 3.3 Nonlinear flow modes and mode coupling or mixing\\
\hspace*{0.9cm} 3.3.1 Connections between seemingly unrelated observables\\
4. Collectivity\\
\hspace*{0.3cm} 4.1 Origin of collectivity\\
\hspace*{0.9cm} 4.1.1 Hydrodynamics\\
\hspace*{0.9cm} 4.1.2 Anisotropic parton-escape models\\
\hspace*{0.9cm} 4.1.3 CGC effective field theory\\
5. Conclusions\\
6. Acknowledgements\\
7. Appendix A\\
\hspace*{0.3cm} 7.1 Moments and cumulants of a probability distribution\\
\hspace*{0.3cm} 7.2 Moment-generating function $M(t)$\\
\hspace*{0.3cm} 7.3 Cumulant-generating function $K(t)$\\
8. Appendix B\\
\hspace*{0.3cm} 8.1 Correlation functions and cumulants\\
9. Appendix C\\
\hspace*{0.3cm} 9.1 Gaussian or normal distribution in 1D\\
\hspace*{0.3cm} 9.2 Gaussian or normal distribution in 2D\\
References\\


\section{Introduction}
\label{intro}
The physics of heavy-ion collisions or even of relativistic heavy-ion
collisions has a long history, see e.g.
\cite{Chin:1978gj,Yano:1978gk}. However, the modern era in this field
began with the advent of the Relativistic Heavy-Ion Collider (RHIC) at
BNL, USA in 2000. The field was further enriched when the Large Hadron
Collider (LHC) at CERN became operational in 2010. The big ideas
driving this field are: to test the predictions of the nonperturbative
Quantum Chromodynamics (QCD), to study the equilibrium and
nonequilibrium (transport) properties of the quark-gluon plasma (QGP),
to map out the QCD phase diagram qualitatively as well as
quantitatively, etc. The relativistic nucleus-nucleus collisions (and
arguably even smaller systems being studied at RHIC and LHC) constitute
the only available tool to produce QGP in the laboratory.

This pedagogical review is restricted to the issue of collectivity as
seen in the large and small systems formed in the nucleus-nucleus,
hadron-nucleus or hadron-hadron collisions at high energies. The
emphasis will be on explaining the mathematical framework underlying
the discussion of the various aspects of the collective flow, which
although quite straightforward, might appear intriguing to the
uninitiated. Although attempt will be made to display the relevant
experimental data, this is not a comprehensive review of the
phenomenology of the wealth of data on collectivity accumulated over
the years.

In the next section we present the modern flow picture. Flow
observables with mixed harmonics are discussed in Sec. 3. Collectivity
in small and large systems is discussed in Sec. 4, which is followed
by Conclusions in Sec. 5. Mathematical preliminaries are presented in
Appendices A to C. Reader unfamiliar with these is advised to go
through the appendices before reading Sec. 2.


\section{Modern Flow Picture}
\label{sec:1}

\begin{figure}
\begin{center}
\resizebox{0.7\columnwidth}{!}{\includegraphics{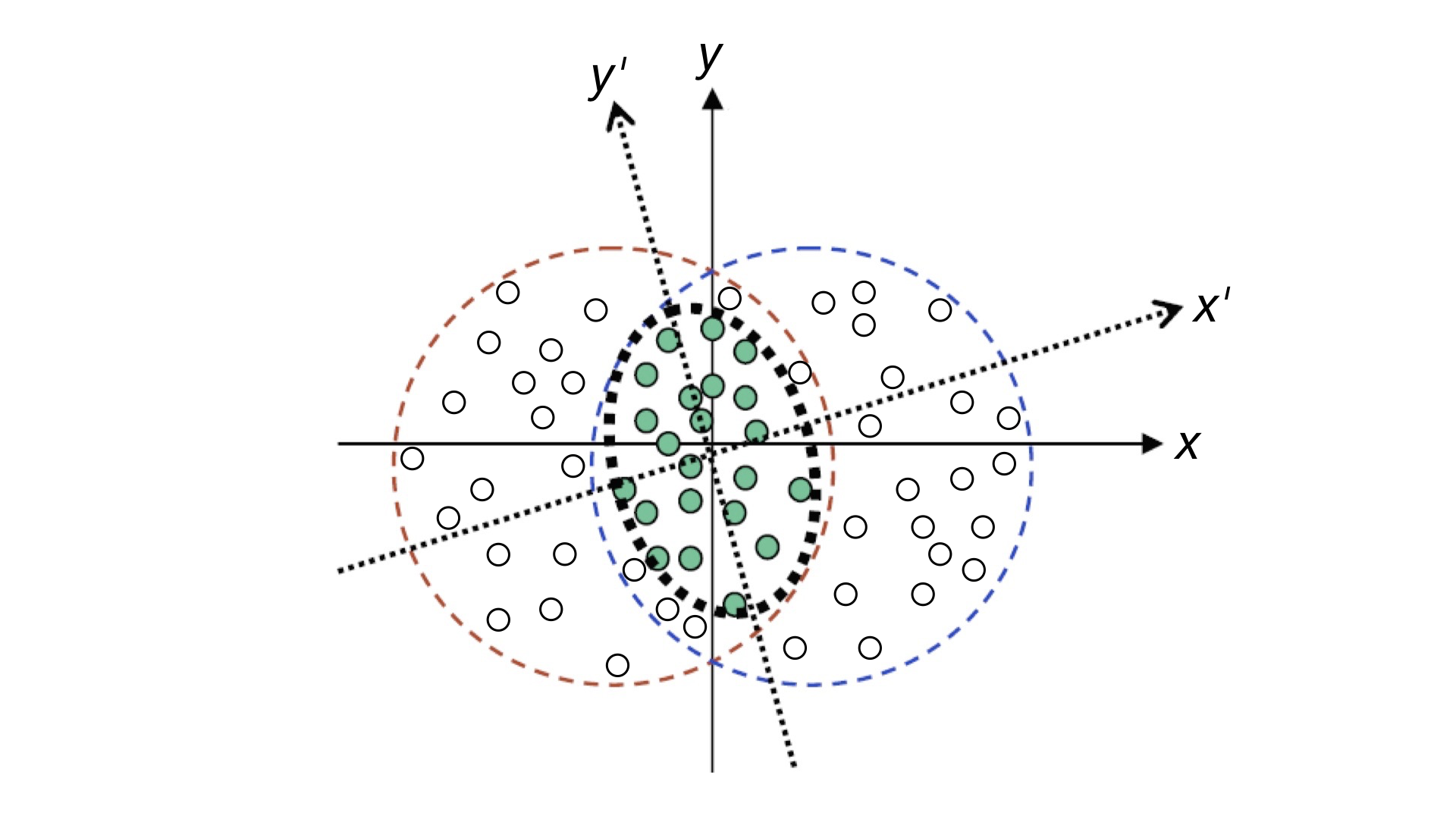}}
\caption{(colour online) Geometry of the collision. Filled circles
  denote the participants and open circles the spectators. $x'$ and
  $y'$ are the principal axes of inertia of the participant
  zone. Figure adapted from \cite{Voloshin:2008dg}.}
\label{fluct}
\end{center}
\end{figure}

We begin by describing the geometry of the collision (of two identical
spherical nuclei) and the various planes that go with it. Two nuclei
are taken to approach each other parallel to the $z$ (or longitudinal
or beam) axis. The origin is taken at the midpoint of the impact
parameter vector. The standard convention for the $x$ and $y$ axes is
as shown in Fig. \ref{fluct}. Due to quantum fluctuations in the wave
functions of the incoming nuclei, the participant zone may be shifted
and/or tilted with respect to the $xy$ frame. $x',y',z'$ are the
principal axes of inertia of the participant zone. $x'$, in
particular, is the direction of the short axis of the participating
nucleon distribution.

\begin{itemize}
\item Reaction Plane (RP): $xz$ plane. It is the plane determined by
  the impact parameter vector and the beam axis. Orientation of the
  reaction plane is not directly measurable.

\item Participant Plane (PP): $x'z'$ plane. Same as the RP if there
  are no initial-state fluctuations. Participant plane is also not
  directly measurable.

\item Event Plane (EP): Estimate of the PP, obtained by measuring the
  direction of maximum final-state particle density. If there are no
  fluctuations, EP is an estimate of RP.\footnote{See
    Eqs. (\ref{eccn1}) and (\ref{Pphi1}) for definitions of PP
    ($\Phi_n$) and EP ($\Psi_n$) specific to harmonic $n$.} 

\end{itemize}

It is established by hydrodynamic calculations and is also expected
naturally that the initial state of collision governs the final state
on an event-by-event basis. In the next two subsections, we describe
how the geometries of the initial and final states, in the coordinate
and momentum spaces respectively, are captured in terms of a few
parameters. One then expects the empirically extracted parameters of
the final state (``flow coefficients'') to depend on the theoretically
calculated parameters of the initial state (``eccentricities''), in a
hopefully simple way.


\subsection{Characterization of the initial state}

Recall the old definition of the eccentricity $\varepsilon_2 \equiv
\mean{y^2-x^2}/ \mean{y^2+x^2}$, which is real. In modern parlance,
the complex eccentricity vector $\mathcal{E}_n$ is defined
as\footnote{$n=1$ forms a special case \cite{Heinz:2013th}.}
\cite{Alver:2010gr,Heinz:2013th}
\begin{equation}
\label{eccn1}
\mathcal{E}_n \equiv \varepsilon_n e^{i n \Phi_n} \equiv 
-\frac{\mean{z^n}}{\mean{|z|^n}}=
-\frac{\mean{r^n e^{i n\varphi}}}{\mean{r^n}}=-\frac{\int z^n e(z,\tau_0)r \,
 dr \, d\varphi}{\int |z|^n e(z,\tau_0)r \, dr \, d\varphi},~~n \ge2,
\end{equation}
where $\varepsilon_n$ is the magnitude and $\Phi_n$ determines the
phase of the complex vector $\mathcal{E}_n$, the complex variable
$z\equiv x+iy=re^{i\varphi}$ refers to a point in the transverse or $xy$ plane, 
$e(z,\tau_0)$ is the
energy (or entropy) density at midrapidity at the initial time
$\tau_0$, i.e., shortly after the collision and $\mean{\cdots}$
denotes the $e(z,\tau_0)$-weighted average over the transverse plane
in a single event after centering: $\mean{z}=\mean{r e^{i
    \varphi}}=0$. Note that $0 \le \varepsilon_n\le 1$ by definition.
$\Phi_n$ is called the participant-plane angle in the $n$-th harmonic.
$\varepsilon_n$ and $\Phi_n$ are given by
\begin{eqnarray} \label{eccn2} \left.
\begin{aligned}
\varepsilon_n =& \frac{\sqrt{\mean{r^n \cos n\varphi}^2+
\mean{r^n \sin n \varphi}^2}}{\mean{r^n}} \, , \\
\Phi_n=&\frac{1}{n}\tan^{-1}\frac{\mean{r^n \sin n\varphi}}
{\mean{r^n \cos n\varphi}} \,,
\end{aligned}\right.
\end{eqnarray}
modulo $2\pi/n$, because the $n$-th order harmonic has an $n$-fold
symmetry in azimuth.
It is easy to check that for $n=2$, with $\Phi_2$ oriented along the
$x$ axis, both Eqs. (\ref{eccn1}) and
(\ref{eccn2}) yield the old definition of the eccentricity
$\varepsilon_2 \equiv \mean{y^2-x^2}/ \mean{y^2+x^2}$. Eccentricity
$\varepsilon_n$ cannot be determined experimentally. However, given a
model for the initial state (e.g., Glauber model), $\varepsilon_n$ can
be calculated from the positions ($r,\varphi$) of the participating
nucleons (or partons) in the transverse plane. The magnitude
$\varepsilon_n$ and the phase $\Phi_n$ fluctuate from event to event.

It is sometimes useful to replace the above definition of $\E_n$ based
on moments by a definition based on cumulants, because cumulants
subtract off the lower-order correlations and retain only the
irreducible correlations \cite{Teaney:2013dta}. The spatial azimuthal
anisotropies then become:
\begin{eqnarray} \label{eccn3} \left.
\begin{aligned}
   \E_2 =& -\frac{\llangle z^2 \rrangle}
{\llangle r^2 \rrangle } \, , \\
   \E_3 =& -\frac{\llangle z^3 \rrangle}
{\llangle r^3 \rrangle } \, , \\
   \E_4 =&- \frac{1}{\llangle r^4 \rrangle} 
\left[ \llangle z^4 \rrangle - 3 \llangle z^2 \rrangle^2 \right] \, , \\
   \E_5 =&- \frac{1}{\llangle r^5 \rrangle} 
\left[ \llangle z^5 \rrangle - 10 \llangle z^2 \rrangle \llangle z^3 \rrangle 
\right] \, ,  \\
   \E_6  =&  -\frac{1}{\llangle r^6 \rrangle }  
\left[ \llangle z^6 \rrangle - 15\llangle z^4 \rrangle \llangle z^2 \rrangle  
- 10 \llangle z^3 \rrangle^2 + 30 \llangle z^2 \rrangle^3  \right] \, .
\end{aligned}\right.
\end{eqnarray}
These will be needed later in Sec. 3.3.
\bigskip

\noindent\fbox{
\parbox{\textwidth}{ {\bf Exercise 1:} Understand Eqs. (\ref{eccn3})
  in the light of the expressions of cumulants given in Appendix A.}}
 

\subsection{Characterization of the final state}

Anisotropy of the azimuthal distribution of the final-state particles
was proposed as a signature of the transverse collective flow
in Ref. \cite{Ollitrault:1992bk}. Characterization of the anisotropy by a 
Fourier decomposition was proposed in Ref. \cite{Voloshin:1994mz}. 
The modern flow picture is a model in which particles in the final
state are emitted randomly and independently according to some
underlying single-particle probability distribution $P(\phi)$ that fluctuates from
event to event \cite{Luzum:2011mm}:
\begin{equation}
\label{Pphi1}
P(\phi)=\frac{1}{2\pi}\sum_{n=-\infty}^{\infty}V_n e^{-in\phi}
=\frac{1}{2\pi}\sum_{n=-\infty}^{\infty}v_n e^{-in(\phi-\Psi_n)},
\end{equation}
where $\phi$ is the azimuthal angle of the momentum of the outgoing
particle and $V_n$ is a complex Fourier flow
coefficient\footnote{$V_n$ is sometimes called a flow vector, and `flow'
and `azimuthal anisotropy' are often used synonymously.} whose
magnitude ($v_n$) and phase ($\Psi_n$) fluctuate from event to event. The
angle $\Psi_n$ is called the event-plane angle or symmetry-plane angle
or reference angle for harmonic $n$. It is easy to check that 
\begin{equation}
\label{Vn}
V_n \equiv v_n \exp(in\Psi_n)=\mean{\exp(in\phi)}, 
\end{equation}
where $\mean{\cdots}$ denotes an average over the probability density
in a single event. As $v_n$ is real, this also means $v_n=\mean{\cos
  n(\phi-\Psi_n)}$ and $\mean{\sin n(\phi-\Psi_n)}=0$. The sine term
vanishes because of the symmetry with respect to the event plane. 
The event-plane angle
$\Psi_n$ can be determined using Eq. (\ref{Vn}):
\begin{equation}
\label{Psin1}
\Psi_n=\frac{1}{n}\tan^{-1}\frac{\mean{\sin n\phi}}{\mean{\cos n\phi}},
~~~(\rm modulo ~ 2\pi/n).
\end{equation}
Experimentally, the flow vector $Q_n$ defined as $Q_n \equiv \sum_i
w_i \exp(i n \phi_i)/\sum_i w_i$ provides an estimate of the vector
$V_n$.  Here $\phi_i$ are the azimuthal angles of the momenta of the detected
particles and $w_i$ are the weights that are generally inserted to optimize the
estimate by accounting for detector nonuniformity and tracking
inefficiency. Experimentally, $\Psi_n$ is estimated as \cite{Poskanzer:1998yz}:
\begin{equation}
\label{Psin2}
\Psi_n=\frac{1}{n}\tan^{-1}\frac{\sum_i w_i \sin n\phi_i}{\sum_i w_i \cos n\phi_i},
~~~(\rm modulo ~ 2\pi/n).
\end{equation}
\bigskip

\noindent\fbox{\parbox{\textwidth}{ {\bf Exercise 2}: Recall that the
    event plane is an estimate of the participant plane. Because the
    number of particles emitted in an event is finite, the
    event-plane angle $\Psi_n$ determined from Eq. (\ref{Psin2})
    fluctuates from event to event around the participant-plane angle
    $\Phi_n$. Hence the true flow $v_n$ would differ from the observed
    flow $v_n^{\rm obs}$.  Show that $v_n=v_n^{\rm obs}/R_n$, where
    the `resolution factor' $R_n$ is given by $R_n=\mean{\cos
      n(\Psi_n-\Phi_n)}< 1$ \cite{Ollitrault:1997di}.}}

\bigskip

Defining $V_{-n}=V_n^*$ (equivalently, $v_n=v_{-n}$, $\Psi_n=\Psi_{-n}$) and using
$V_0=v_0=1$, one can write $P(\phi)$ also as
\begin{equation}
\label{Pphi2}
P(\phi)= \frac{1}{2\pi} \left( 1+ 2\sum_{n=1}^\infty  v_n \cos
n(\phi-\Psi_n) \right),
\end{equation}
which is real. As a result, in the final state, the azimuthal
distribution of particles that fluctuates event to event is
\begin{equation}
\label{dNdphi}
\frac{dN}{d\phi}=\frac{N}{2\pi} \left( 1+ 2\sum_{n=1}^\infty  v_n \cos
n(\phi-\Psi_n) \right).
\end{equation}
The angle-independent leading term is called the radial flow.
The coefficients $v_n$ parametrize the momentum anisotropy of the
final-state particle yield. They go by the names directed or dipolar
($n=1$), elliptic ($n=2$), triangular ($n=3$), quadrangular ($n=4$),
pentagonal ($n=5$), $\cdots$ flows.\footnote{These names are
  suggestive of the shapes of the polar plots $r=1+2v_n \cos n\phi$,
  for $0 < v_n \ll 1$.} Recently, ALICE collaboration has presented results
on flow harmonics up to $n=9$ \cite{Acharya:2020taj}.

In general, $v_n$ and $\Psi_n$ depend on the transverse momentum
$(p_T)$ and pseudorapidity $(\eta)$. Thus the event-plane angle is not
a unique angle for the entire event. It is so only for nonfluctuating
smooth initial conditions. The $p_T$ dependence of $\Psi_n$ may arise
due to differently-oriented and different-sized hot spots in the
transverse plane emitting particles in different directions with
different $\mean{p_T}$ \cite{Khachatryan:2015oea}. 
The $\eta$ dependence of $\Psi_n$ may arise
due to a fireball that is twisted or torqued in the longitudinal
direction \cite{Bozek:2010vz}.

\medskip

Note that the $\varepsilon_n$ coefficients characterize the various shape
components of the fluctuating {\it initial} profile in the {\it
  coordinate} space, just as the $v_n$ coefficients characterize those of
the fluctuating {\it final} state in the {\it momentum} space.


\subsection{Flow measurements}

Theoretically, 
\begin{equation}
V_n (p_T,\eta) \equiv v_n e^{in \Psi_n} (p_T,\eta) = \frac
{\int d\phi e^{i n \phi}\frac{dN}{p_T dp_T d\eta d\phi}}
{\int d\phi             \frac{dN}{p_T dp_T d\eta d\phi}}
= \mean{e^{i n \phi}},
\end{equation}
where $dN/p_T dp_T d\eta d\phi$ is the particle distribution ---
usually charged hadron distribution --- in an event. 
$V_n(p_T)$ and $V_n(\eta)$ can be
defined similarly, by carrying out integrations over $\eta$ or $p_T$,
respectively, over appropriate ranges, besides the integration over
$\phi$. Fully integrated flow $V_n$ results from the integration over
both $p_T$ and $\eta$.

Since the orientations of the reaction plane and the participant plane are
unknown, flow $v_n$ is usually measured experimentally from
event-averaged azimuthal correlations between outgoing
particles. Determination of the flow coefficients using multiparticle
correlations proceeds as follows. Two-, four-. six- and eight-particle
azimuthal correlations are defined as
\begin{eqnarray}
\label{mpc1}
  \left.\begin{aligned}
\langle\langle 2 \rangle\rangle &= \langle\langle e^{in(\phi_{1}-\phi_{2})} 
\rangle\rangle, \\
\langle\langle 4 \rangle\rangle &= \langle\langle e^{in(\phi_{1}+\phi_{2}-
\phi_{3}-\phi_{4})} \rangle\rangle, \\
\langle\langle 6 \rangle\rangle &= \langle\langle e^{in(\phi_{1}+\phi_{2}+\phi_{3}
-\phi_{4}-\phi_{5}-\phi_{6})} \rangle\rangle, \\
\langle\langle 8 \rangle\rangle &= \langle\langle e^{in(\phi_{1}+\phi_{2}+\phi_{3}
+\phi_{4}-\phi_{5}-\phi_{6}-\phi_{7}-\phi_{8})} \rangle\rangle,
\end{aligned}\right.
\end{eqnarray}
where $\langle\langle \cdots \rangle\rangle$ denotes averaging over
all multiplets in a single collision event and then over all events in
a given centrality class. Note that all these `observables' are
invariant under rotation in the azimuthal plane, as they should be.
(As a counterexample, consider $\langle\langle
e^{in(\phi_{1}+\phi_{2}-\phi_3)} \rangle\rangle$. To make it
invariant, we may replace $\phi_3$ by $2\phi_3$, which, however, mixes
harmonics $n$ and $2n$. Mixed-harmonic observables will be discussed
in Sec. 3.) Multiparticle cumulants are defined as 
\cite{Borghini:2001vi}
\begin{eqnarray}
\label{mpc2}
    \left.\begin{aligned}
      c_{n}\{2\} &= \langle\langle 2 \rangle\rangle, \\
      c_{n}\{4\} &= \langle\langle 4 \rangle\rangle - 2 \langle\langle 2 
\rangle\rangle^{2}, \\
      c_{n}\{6\} &= \langle\langle 6 \rangle\rangle - 9 \langle\langle 2 
\rangle\rangle \langle\langle 4 \rangle\rangle + 12 \langle\langle 2 
\rangle\rangle^{3}, \\
      c_{n}\{8\} &= \langle\langle 8 \rangle\rangle - 16 \langle\langle 2 
\rangle\rangle \langle\langle 6 \rangle\rangle - 18 \langle\langle 4 
\rangle\rangle^{2}
      + 144 \langle\langle 2 \rangle\rangle^{2}  \langle\langle 4 \rangle\rangle
- 144 \langle\langle 2 \rangle\rangle^{4}.
      \end{aligned}\right.
\end{eqnarray}

\bigskip

\noindent\fbox{
\parbox{\textwidth}{
{\bf Exercise 3:} Understand Eqs. (\ref{mpc2}) in the light of the definitions
of cumulants given in Eq. (B2) of Appendix B.
}}

\bigskip

The two-particle correlation can be rewritten as
\begin{eqnarray}\label{mpc3}\left.\begin{aligned}
\langle\langle 2 \rangle\rangle =& \langle\langle e^{in(\phi_{1}-\phi_{2})} 
\rangle\rangle 
= \langle\langle e^{in(\phi_{1}-\Psi_n)}e^{-in(\phi_{2}-\Psi_n)} \rangle\rangle \\
=& \langle\langle e^{in(\phi_{1}-\Psi_n)}\rangle
\langle e^{-in(\phi_{2}-\Psi_n)} \rangle\rangle =\mean{v_n^2},
\end{aligned}\right.\end{eqnarray}
where in the first line we assumed that $\Psi_n$ is a global phase
angle for all the particles selected for averaging, and in the second
line we assumed that there are no nonflow
correlations\footnote{Nonflow correlations are not related to the
initial-state geometry and hence not associated with the
  symmetry plane $\Psi_n$, but arise due to jets, particle decays,
  etc. They are of short range.}, or equivalently, that the angles $(\phi_{1}-\Psi_n)$ and
$(\phi_{2}-\Psi_n)$ are statistically independent. Note that any
dependence on the symmetry plane is eliminated by construction.  The
higher-order correlations $\langle\langle 4 \rangle\rangle$,
$\langle\langle 6 \rangle\rangle$ and $\langle\langle 8
\rangle\rangle$ can be treated similarly. Thus the multiparticle 
cumulants take the form:
\begin{eqnarray}
\label{mpc4}
    \left.\begin{aligned}
      c_{n}\{2\} &= \langle v_n^2\rangle, \\
      c_{n}\{4\} &= \langle v_n^4\rangle - 2 \langle v_n^2 \rangle^{2}, \\
      c_{n}\{6\} &= \langle v_n^6 \rangle - 9 \langle v_n^2 
\rangle \langle v_n^4 \rangle + 12 \langle v_n^2 
\rangle^{3}, \\
      c_{n}\{8\} &= \langle v_n^8 \rangle - 16 \langle v_n^2 
\rangle \langle v_n^6 \rangle - 18 \langle v_n^4 
\rangle^{2}
      + 144 \langle v_n^2 \rangle^{2} \langle v_n^4 \rangle
- 144 \langle v_n^2 \rangle^{4}.
      \end{aligned}\right.
\end{eqnarray}

Finally, the flow coefficients are given by\footnote{$v_n\{m\}$ are
  also called multiparticle cumulants of order $m$ of the flow $v_n$.}
\begin{eqnarray}
\label{mpc5}
\left.\begin{aligned}
v_n\{2\} &= \sqrt{c_n\{2\}}= \sqrt{\langle v_n^{2} \rangle}, \\
v_n\{4\} &= \sqrt[4]{- c_n\{4\}}
= \sqrt[4]{2 \langle v_n^{2} \rangle^2 - \langle v_n^{4} \rangle}, \\
v_n\{6\} &= \sqrt[6]{\frac{1}{4} c_n\{6\}}
= \sqrt[6]{\frac{1}{4}(\langle v_n^{6} \rangle - 9 \langle v_n^{2} \rangle \langle 
v_n^{4} \rangle + 12 \langle v_n^{2} \rangle^3}), \\
v_n\{8\} &= \sqrt[8]{-\frac{1}{33} c_n\{8\}} \\
&= \sqrt[8]{-\frac{1}{33}(\langle v_n^{8} \rangle - 16 \langle v_n^{2} \rangle
 \langle v_
n^{6} \rangle - 18 \langle v_n^{4} \rangle^2 + 144 \langle v_n^{2} \rangle^2 
\langle
 v_n^{4} \rangle - 144 \langle v_n^{2} \rangle^4}).
\end{aligned}\right.
\end{eqnarray}
The coefficients $-1$, 1/4 and $-1/33$ appearing in front of
$c_n\{4\}$, $c_n\{6\}$ and $c_n\{8\}$, respectively, in Eqs. (\ref{mpc5}), can be
understood as follows: If the magnitude of the flow vector does not
fluctuate event-to-event, then $\mean{v_n^k}=v_n^k$, and 
it is clear from Eq. (\ref{mpc4}) that
$c_n\{2\}=v_n^2 (> 0)$, $c_n\{4\}=-v_n^4 (< 0)$, $c_n\{6\}=4v_n^6 (> 0)$,
$c_n\{8\}=-33v_n^8 (< 0)$. These when substituted in Eq. (\ref{mpc5}) ensure
that $v_n\{2\}=v_n\{4\}=v_n\{6\}=v_n\{8\}=v_n$, as expected.
An advantage of multiparticle cumulants $v_n\{m\}$ is that they
suppress nonflow contribution.

For the sake of completeness, we present the inverse of Eqs. (\ref{mpc5}):
\begin{eqnarray}\label{}\left.\begin{aligned}
\langle v_n^2\rangle &= v_{n}\{2\}^2,\\
\langle v_n^4\rangle &= -v_{n}\{4\}^4 + 2v_{n}\{2\}^4,\\
\langle v_n^6\rangle &= 4v_{n}\{6\}^6 -9 v_{n}\{2\}^2 \,v_{n}\{4\}^4+6v_{n}\{2\}^6,\\
\langle v_n^8\rangle &= -33v_{n}\{8\}^8 +64 v_{n}\{6\}^6 \,v_{n}\{2\}^2 
+18v_{n}\{4\}^8 -72 v_{n}\{2\}^4 \,v_{n}\{4\}^4 +24v_{n}\{2\}^8 .
\end{aligned}\right.\end{eqnarray}

The flow coefficients in Eq. (\ref{mpc5}) are driven by the corresponding 
initial-state eccentricities defined as \cite{Bhalerao:2006tp}:
\begin{eqnarray}
\label{mpc6}
\left.\begin{aligned}
\varepsilon_n\{2\} &= \sqrt{\langle \varepsilon_n^{2} \rangle}, \\
\varepsilon_n\{4\} &
= \sqrt[4]{2 \langle \varepsilon_n^{2} \rangle^2 - \langle \varepsilon_n^{4} \rangle}, \\
\varepsilon_n\{6\} &=
\sqrt[6]{\frac{1}{4}(\langle \varepsilon_n^{6} \rangle - 9 \langle \varepsilon_n^{2} \rangle \langle 
\varepsilon_n^{4} \rangle + 12 \langle \varepsilon_n^{2} \rangle^3}), \\
\varepsilon_n\{8\}
&= \sqrt[8]{-\frac{1}{33}(\langle \varepsilon_n^{8} \rangle - 16 \langle \varepsilon_n^{2} \rangle
 \langle \varepsilon_n^{6} \rangle - 18 \langle \varepsilon_n^{4} \rangle^2 + 144 \langle \varepsilon_n^{2} \rangle^2 
\langle
 \varepsilon_n^{4} \rangle - 144 \langle \varepsilon_n^{2} \rangle^4}).
\end{aligned}\right.
\end{eqnarray}

\medskip

Consider $V_2=v_2 \exp(i\,2\, \Psi_2), \, V_4=v_4 \exp(i\,4\, \Psi_4)$ and
$V_6=v_6 \exp(i\,6\, \Psi_6)$. In the foregoing discussion, we implicitly
assumed that $V_4$ and $V_6$ are to be analysed with respect to their
respective reference angles, $\Psi_4$ and $\Psi_6$. But that is not
necessary. We could alternately analyse $V_4$ with respect to the
direction of $\Psi_2$, leading to $\exp(i4(\Psi_4-\Psi_2))$.
Similarly, analysing $V_6$ with respect to the direction of $\Psi_2$
or of $\Psi_3$ would involve $\exp(i6 (\Psi_6-\Psi_2))$ or $\exp(i6
(\Psi_6-\Psi_3))$. We shall encounter such quantities later when we
discuss event-plane correlators (Sec. 3.2).

For recent theoretical developments about calculating and analyzing
multiparticle cumulants to arbitrary order, see
\cite{DiFrancesco:2016srj,Moravcova:2020wnf}.


\subsection{Probability density function (PDF)}

\begin{figure}
\resizebox{0.99\columnwidth}{!}{\includegraphics{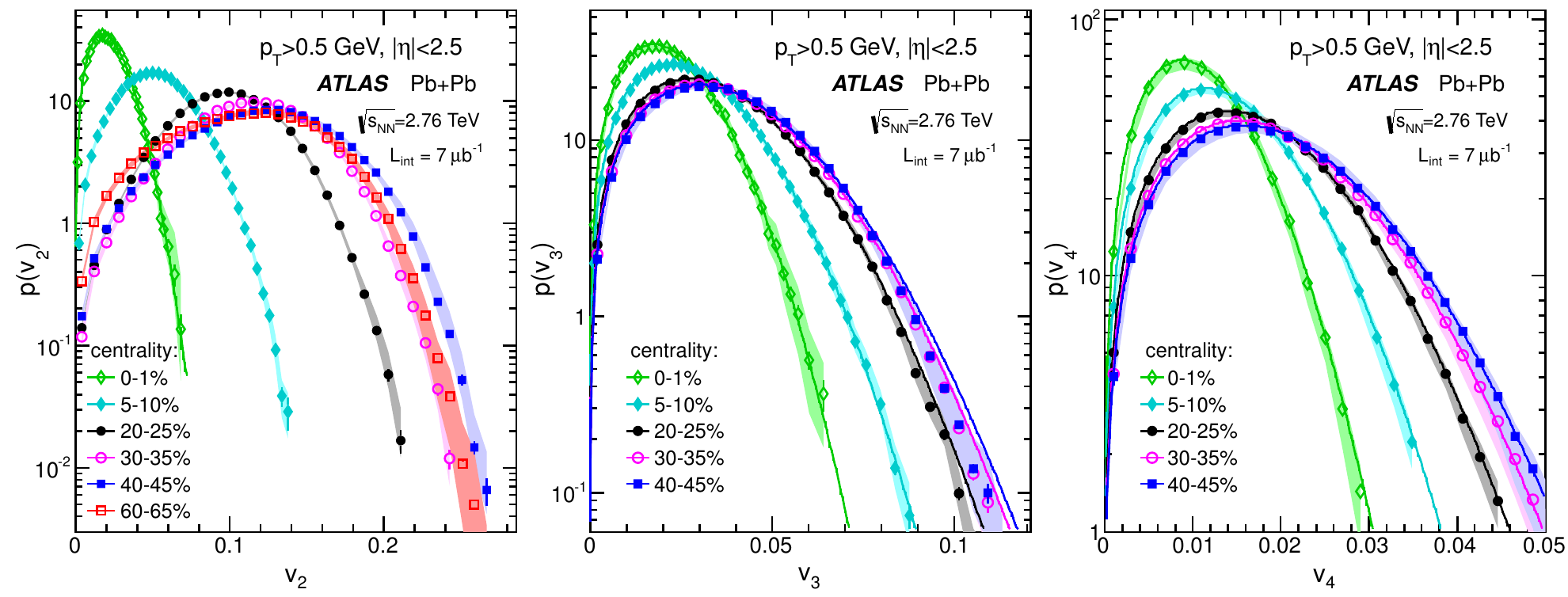}}
\caption{(colour online) ATLAS data on probability density function
  $P(v_n)$ in several centrality intervals for n = 2 (left panel), n =
  3 (middle panel) and n = 4 (right panel) \cite{Aad:2013xma}.}
\label{Pvn}
\end{figure}

As stated above, the flow magnitude ($v_n$) and phase ($\Psi_n$) (or
equivalently, $x$ and $y$ components of $V_n$) fluctuate event to
event. The probability density function $P(v_n)$ has been measured.
Figure \ref{Pvn} shows the ATLAS data on $P(v_2),\, P(v_3)$ and $P(v_4)$
for various centrality bins at $\sqrt{s_{\rm NN}}=2.76$ TeV
\cite{Aad:2013xma}. As one goes from central to peripheral collisions,
the distributions broaden reflecting the gradual increase of $v_n$
with centrality. Centrality dependence of $P(v_2)$ is stronger than
that of $P(v_3)$ and $P(v_4)$, because $\varepsilon_2$ which drives
$v_2$ changes more dramatically from central to peripheral collisions
than do $\varepsilon_3$ and $\varepsilon_4$. Similar results were
reported by ALICE at $\sqrt{s_{\rm NN}}=5.02$ TeV
\cite{Acharya:2018lmh}. Flow fluctuations affect two- and
multiparticle cumulants in different ways (Exercise 4).

\bigskip

\noindent\fbox{
\parbox{\textwidth}{ {\bf Exercise 4}: Define $\sigma_v^2 \equiv
  \mean{v^2}-\mean{v}^2$. (Harmonic index $n$ is suppressed for
  simplicity of notation.) Note that $v\{2\}$, $v\{4\}$ and $v\{6\}$
  defined in Eq. (\ref{mpc5}) are various functions of $v$, denoted
  generically by $f(v)$. Taylor expand $f(v)$ around the mean flow
  $\mean{v}$ to the second order and take the mean to get
  $\mean{f(v)}=f(\mean{v})+\sigma_v^2 f''(\mean{v})/2$. Show that,
  $v\{2\}^2 = \mean{v}^2+\sigma_v^2$ and $v\{4\}^2 \approx
  \mean{v}^2-\sigma_v^2 \approx v\{6\}^2$, and hence $v\{2\} \approx
  \mean{v}+\sigma_v^2/(2\mean{v})$ and $v\{4\} \approx
  \mean{v}-\sigma_v^2/(2\mean{v}) \approx v\{6\}$. Thus
  $v\{2\}>\mean{v}>v\{4\}\approx v\{6\}$. In other words,
  fluctuations, in general, tend to enhance $v\{2\}$ and suppress
  $v\{4\}$, $v\{6\}$ compared to $\mean{v}$
  \cite{Voloshin:2008dg,Ollitrault:2009ie}. (N.B. Here, we did not assume the
  fluctuations to be Gaussian.)}}

\bigskip

As a measure of the relative flow fluctuations, one defines:
\begin{equation}
\label{Fvn}
F(v_n)\equiv\sqrt{\frac{v_n\{2\}^2-v_n\{4\}^2}{v_n\{2\}^2+v_n\{4\}^2}}
=\frac{\sigma_{v_n}}{\mean{v_n}},
\end{equation}
where the last equality follows from Exercise 4. Note that
Eq. (\ref{Fvn}) is based on two assumptions: (a) there are no nonflow
correlations in $v_n\{2\}$ and $v_n\{4\}$, and (b) $\sigma_{v_n} \ll
\mean{v_n}$.

\begin{figure}
\begin{center}
\resizebox{0.5\columnwidth}{!}{\includegraphics{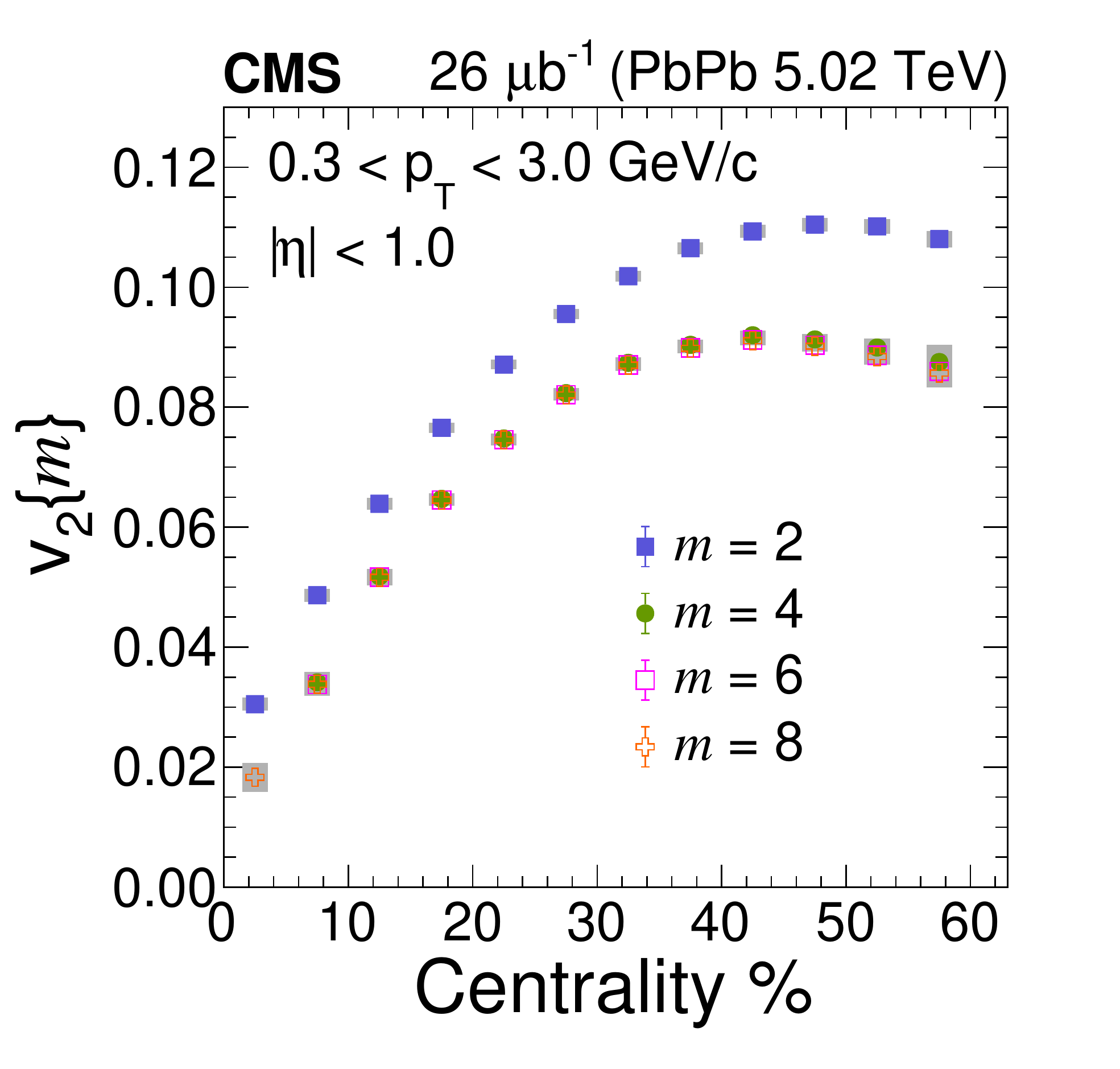}}
\caption{(colour online) Note that $v_2\{2\}>v_2\{4\} \approx v_2\{6\}\approx
  v_2\{8\}$. This is consistent with the Gaussian ansatz
  \cite{Voloshin:2007pc}; see Exercise 5. Figure from
  \cite{Sirunyan:2017fts}.}
\label{v2m}
\end{center}
\end{figure}

Flow coefficients $v_n\{m\}, m=2,4,6,8$ have also been
measured. Consider first $n=2$. Figure \ref{v2m} shows the CMS data
\cite{Sirunyan:2017fts} on $v_2\{m\}$ for $m=2,4,6,8$. Evidently,
$v_2\{2\}>v_2\{4\} \approx v_2\{6\}\approx v_2\{8\}$. It is easy to
understand this feature of the data by making a simple Gaussian ansatz
\cite{Voloshin:2007pc} for the fluctuations of the vector $V_n$ in the
transverse ($xy$) plane. Denoting $x$ and $y$ components of $V_2$
by $v_x$ and $v_y$ for simplicity of notation, one assumes the
probability distribution to be
\begin{equation}
\label{gaussv2}
P(v_x,v_y)=\frac{1}{2\pi \sigma^2} \exp\left(-\frac{(v_x-{\bar
    v})^2+v_y^2}{2\sigma^2} \right),
\end{equation}
where $\bar v = \mean{v_x}$, $\mean{v_y}=0$ and
$\sigma^2=\sigma_x^2=\sigma_y^2$ is the variance of the Gaussian
distribution. Assume further that $v_x$ and $v_y$ are uncorrelated.

\bigskip

\noindent\fbox{
\parbox{\textwidth}{ {\bf Exercise 5}: Using Eqs. (\ref{mpc5}) and
  (\ref{gaussv2}) and the moments of the Gaussian distribution given in
  Table 1 in Appendix C, show that $v_2\{2\}=\sqrt{{\bar v}^2+2\sigma^2}$ and
  $v_2\{4\}= v_2\{6\}=v_2\{8\}={\bar v}$. This shows that the data in
  Fig. \ref{v2m} is consistent with the Gaussian ansatz
  \cite{Voloshin:2007pc}.  } }

\bigskip

\begin{figure}
\resizebox{0.99\columnwidth}{!}{\includegraphics{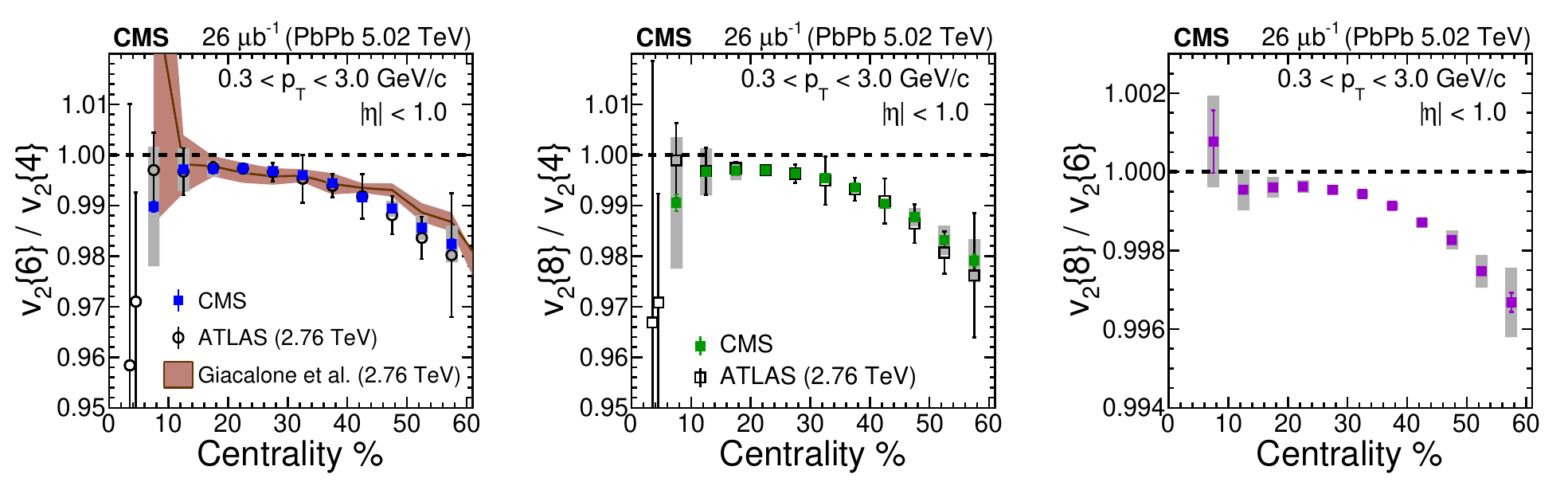}}
\caption{(colour online) Note the small differences in the magnitudes of $v_2\{4\}$,
  $v_2\{6\}$ and $v_2\{8\}$. This is the evidence for non-Gaussianity
  in the elliptic flow fluctuations. Figure from
  \cite{Sirunyan:2017fts}.}
\label{ng2}
\end{figure}

\noindent A more careful inspection of the data, however, shows that
there are small differences in the magnitudes of $v_2\{4\}$,
$v_2\{6\}$ and $v_2\{8\}$, indicating non-Gaussian fluctuations in the
data on $v_2$ (Fig. \ref{ng2}).

We next turn to $n=3$. Unlike $v_2$, which can be nonzero even in the
absence of fluctuations, $v_3$ (at midrapidity) arises only due to
fluctuations. So the Gaussian ansatz for $v_3$ is
\begin{equation}
\label{gaussv3}
P(v_x,v_y)=\frac{1}{2\pi \sigma^2} \exp\left(-\frac{v_x^2
+v_y^2}{2\sigma^2} \right).
\end{equation}

\bigskip

\noindent\fbox{
\parbox{\textwidth}{ {\bf Exercise 6}: Using Eqs. (\ref{mpc5}) and
  (\ref{gaussv3}) and the moments of the Gaussian distribution given in
  Table 1 in Appendix C, show that $v_3\{2\}=\sqrt{2}\sigma$ and
  $v_3\{4\}= v_3\{6\}=v_3\{8\}=0$.}}

\begin{figure}
\begin{center}
\resizebox{0.45\columnwidth}{!}{\includegraphics{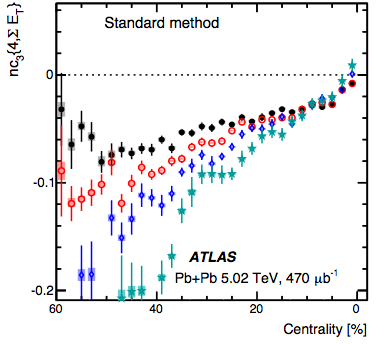}}
\caption{(colour online) $nc_3\{4\} \equiv -v_3\{4\}^4/v_3\{2\}^4$.
  Data points with various colours correspond to different $p_T$
  ranges. Note the nonzero value of $v_3\{4\}$. This is the evidence
  for non-Gaussianity in the triangular flow fluctuations; see
  Exercise 6. Figure from \cite{Aaboud:2019sma}.}
\label{ng3}
\end{center}
\end{figure}

\bigskip
\noindent Figure \ref{ng3} shows ATLAS results for
\begin{equation}
\label{ncn4}
nc_n\{4\} \equiv \frac{c_n\{4\}}{c_n\{2\}^2}=-\frac{v_n\{4\}^4}{v_n\{2\}^4} = \frac{\mean{v_n^4}}{\mean{v_n^2}^2}-2, 
\end{equation}
for $n=3$. It is clear from the figure that $v_3\{4\} \ne 0$, except
for the most central collisions. This is an 
evidence for non-Gaussianity in the triangular flow
fluctuations. Note that $\mean{v_n^4}/\mean{v_n^2}^2-2$ occurring in
Eq. (\ref{ncn4}) is a measure of excess kurtosis; see the discussion of
$f(r)$ in Appendix C.

\bigskip
\noindent\fbox{
\parbox{\textwidth}{ {\bf Exercise 7}: Recall that the cumulants
  (Eq. (\ref{mpc4})) are functions of even moments of the
  distribution of $v_n$. For $n=2$ (the most central collisions only)
  and $n=3$ (all centralities), flow is solely due to fluctuations. If
  the statistics of these fluctuations is a 2-dimensional Gaussian
  (Appendix C), then show that the scaled even moments $m_n^{(k)}$
  satisfy:
\[
m_n^{(k)} \equiv \frac{\mean{v_n^{2k}}}{\mean{v_n^{2(k-1)}}\mean{v_n^2}} = k,
~~~~~(k=2,3,4).
\]
These expectations are borne out quite well in AMPT model calculations
\cite{Bhalerao:2014xra}.
}}
\medskip

If the flow fluctuations are not Gaussian, then what is the right
underlying PDF? A related question is what is the PDF for the
fluctuating initial eccentricity? This issue has been discussed in
the literature: A distribution that was proposed early on was the
Bessel-Gaussian distribution \cite{Voloshin:2007pc}. It worked well
for central collisions, but not so well for peripheral collisions. A
new parametrization named the Elliptic Power distribution, for the PDF
of the initial eccentricity at a fixed centrality, was proposed in
Ref. \cite{Yan:2014afa}. Unlike the Bessel-Gaussian distribution, this
fits several Monte Carlo models of the initial state for all
centralities. 

For more recent works that probe the skewness and kurtosis of the
non-Gaussian fluctuations in heavy-ion collisions, see
\cite{Giacalone:2016eyu,Bhalerao:2018anl}. In cosmology, the
primordial non-Gaussianity is consistent with zero, and the
present-day non-Gaussianity is the result of the evolution of the
universe. In heavy-ion collisions, on the other hand, the primordial
non-Gaussianity is sizable, and it is partially washed out as a result
of the system evolution \cite{Bhalerao:2019fzp}. Future
high-statistics measurements of kurtosis of elliptic flow fluctuations
would provide deeper insights into the hydrodynamic behaviour and the
equation of state of quark-gluon plasma.

We discussed in this section the probability density function
$P(v_n)$. Ideally, one would like to measure the full probability
density function $P(V_1,V_2,\cdots,V_n)=\\
P(v_1,v_2,\cdots,v_n,\Psi_1,\Psi_2, \cdots, \Psi_n)$. 
We are far from that goal.


\subsection{Ridge}

\begin{figure}
\centering
\includegraphics[width=0.45\textwidth]
{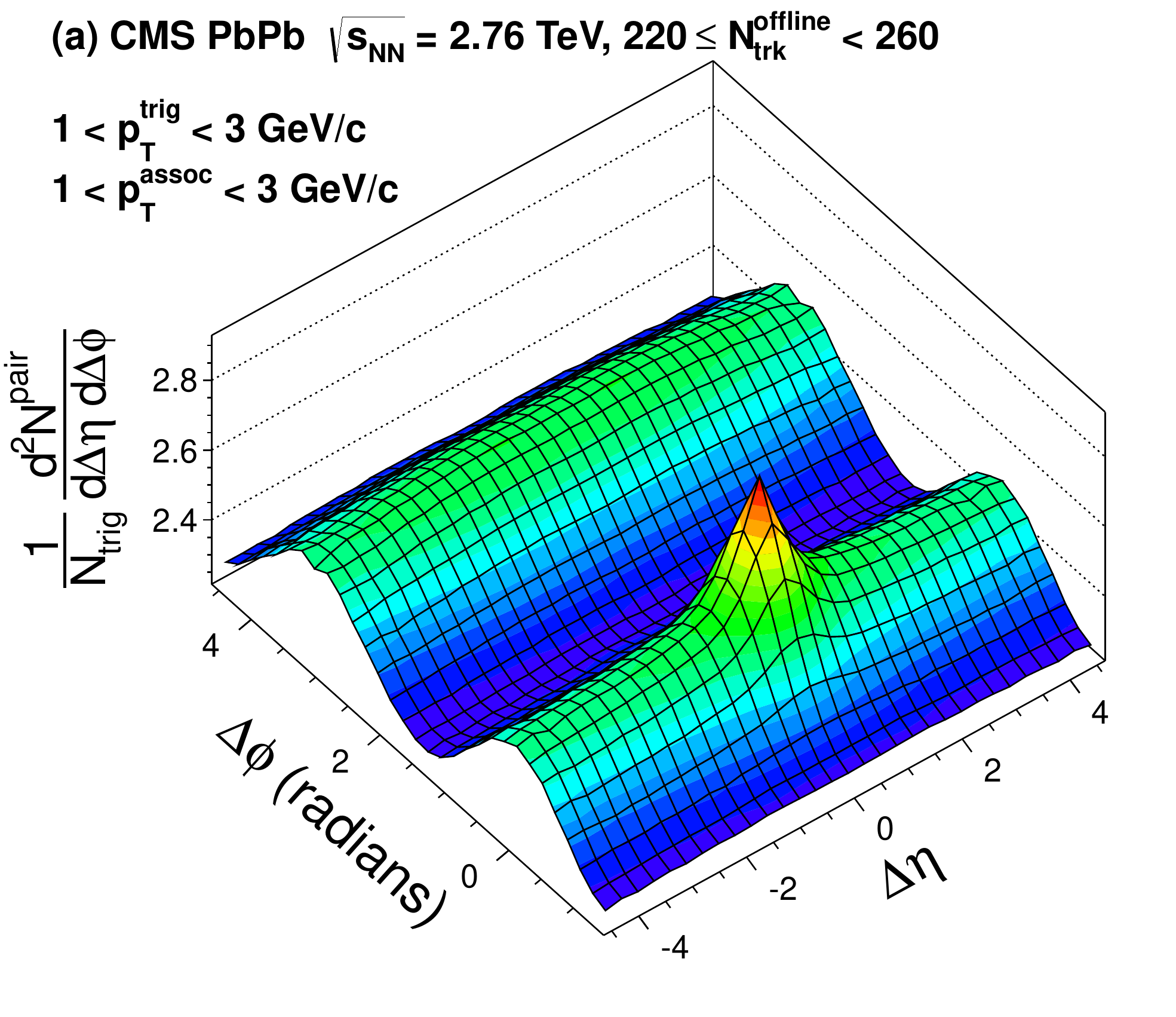}
\includegraphics[width=0.45\textwidth]
{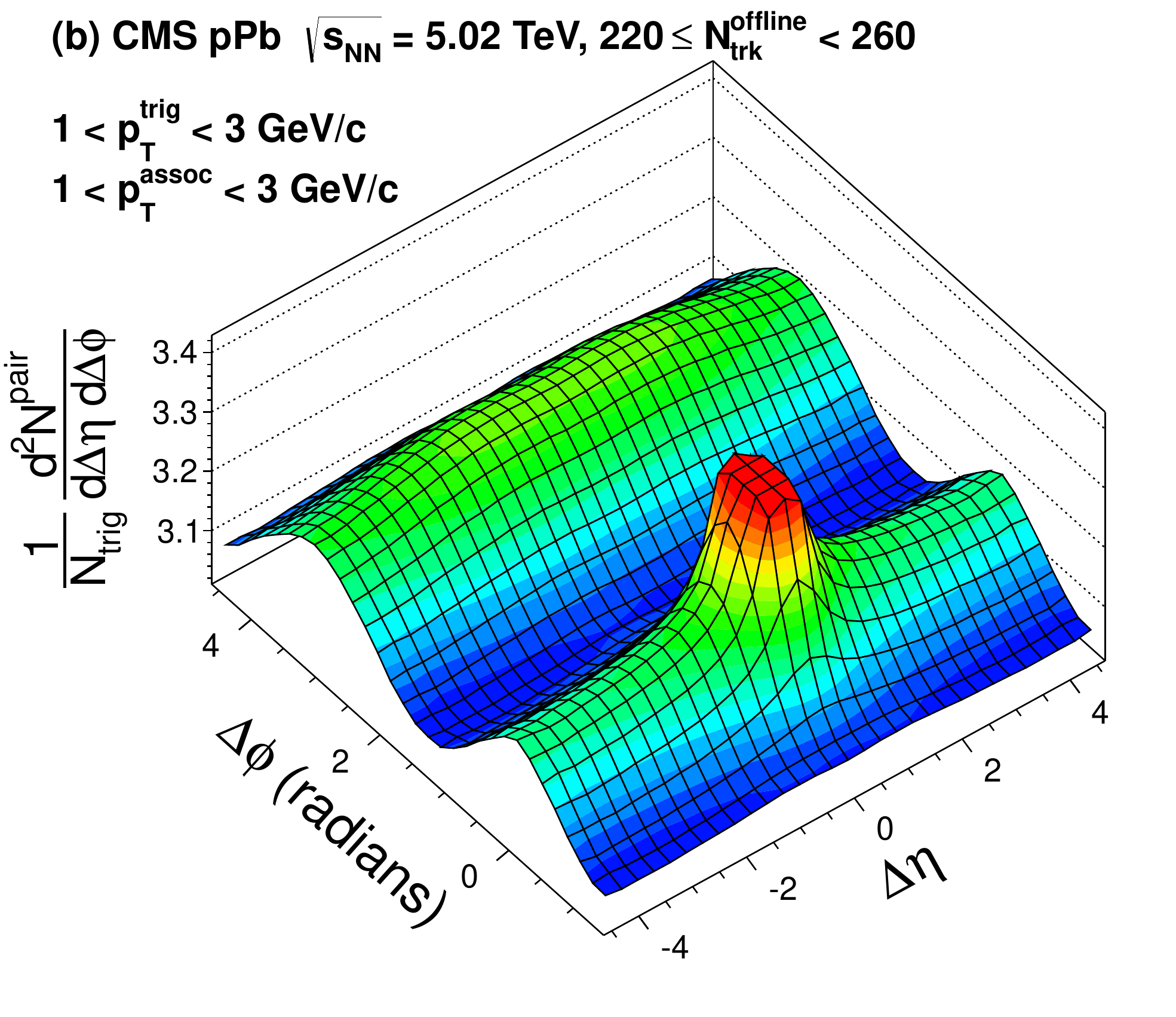}
\caption{(colour online) Two-charged-particle correlations in (a) PbPb and (b) pPb
  collisions observed by CMS \cite{Chatrchyan:2013nka}. Semiperipheral
  PbPb collisions are selected so that particle multiplicities are
  similar to those in the pPb case. In (b), the sharp near-side peak
  from jet correlations is truncated to emphasize the structure
  outside that region.}
\label{ridge1}
\end{figure}

\begin{figure}
\centering
\includegraphics[width=0.45\textwidth]
{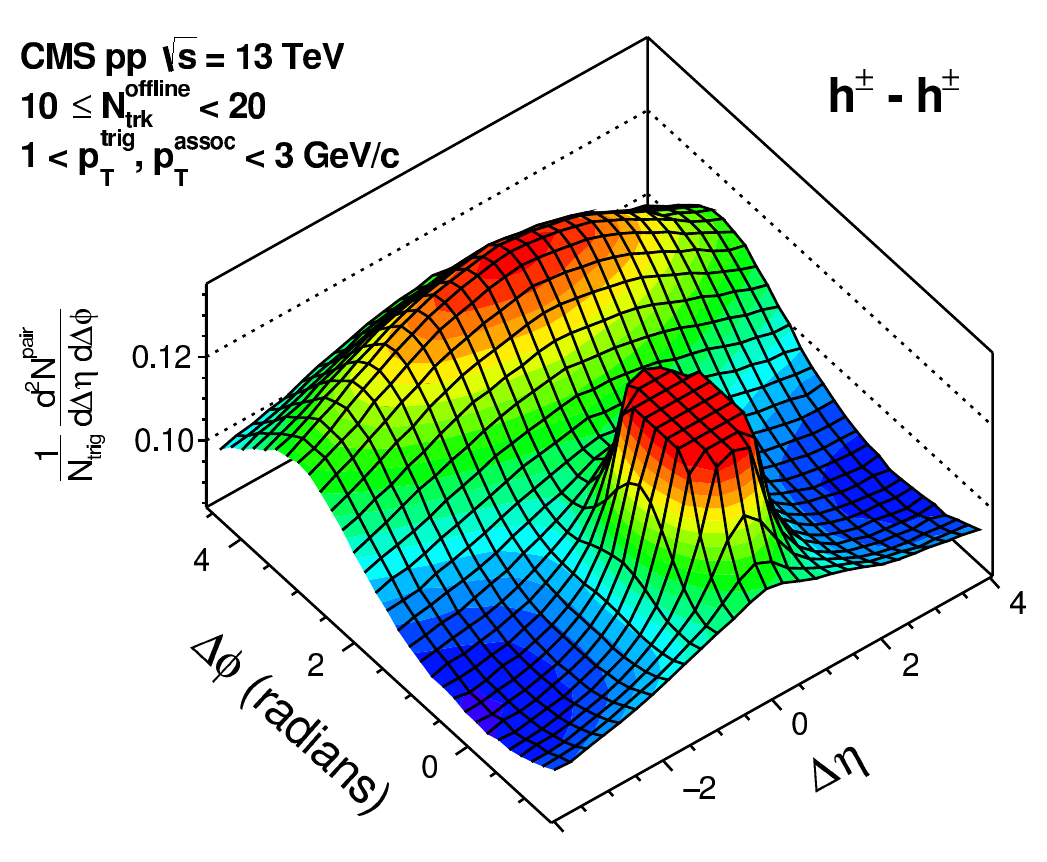}
\includegraphics[width=0.45\textwidth]
{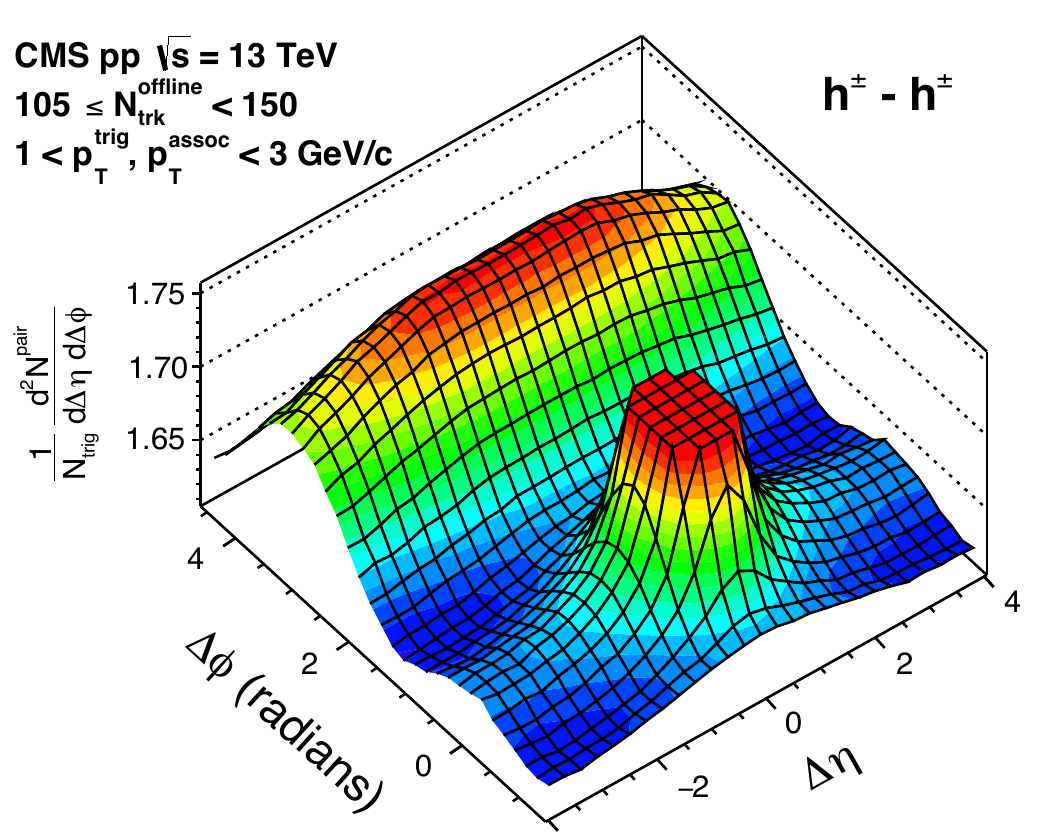}
\caption{(colour online) Two-charged-particle correlations in (left) low-multiplicity (10-20)
  and (right) high-multiplicity (105-150) pp collisions observed by CMS
  \cite{Khachatryan:2016txc}.}
\label{ridge2}
\end{figure}

Equation (\ref{dNdphi}) is the Fourier expansion of the
single-particle distribution in the azimuthal plane. Fourier expansion
of the two-particle distribution is similarly given by
\begin{equation}
\label{dNdDphi}
\frac{dN^{\rm pair}}{d\Delta\phi}=\frac{N^{\rm pair}}{2\pi} \left(1+
2\sum_{n=1}^\infty V_{n\Delta} \cos n\Delta\phi \right),
\end{equation}
where $\Delta \phi = \phi_1-\phi_2$ and
$V_{n\Delta}=\mean{\exp(in\Delta\phi)}$ are the two-particle Fourier
coefficients which fluctuate from event to event. One can similarly
define $d^2N^{\rm pair}/d\Delta \eta \,d\Delta\phi$, where $\Delta
\eta = \eta_1-\eta_2$. This quantity is displayed in
Figs. \ref{ridge1} and \ref{ridge2}.

Figure \ref{ridge1} shows two-charged-particle correlations as a
function of $\Delta \eta$ and $\Delta \phi$, measured by CMS
\cite{Chatrchyan:2013nka}, in PbPb and pPb collisions. The sharp
near-side ($\Delta \phi \approx 0$) peak at $\Delta \eta \approx 0$ is
the signature of short-range correlations arising due to jet
fragmentation. Ignoring this, the striking long-range (large $\Delta
\eta$) wave-like structure that we see on the near-side is called the
ridge. It indicates that particles widely separated in pseudorapidity
experience a collective push in the same azimuthal
direction. Production of a small drop of fluid is a natural
explanation of such a correlation present at all rapidities.  The
peaking around $\Delta \phi = 0$ and $\pi$ is the signature of the
dominant elliptic flow ($v_2$) with some admixture of other harmonics
notably $v_3$ (see Exercise 8). Thus not only the flow harmonics
$v_n$, but even the correlations observed between particles separated
by a large pseudorapidity gap could be explained by a collective
behaviour. Causality requires that the long-range structure seen in
the ridge originates in the earliest stages of the collision when
the transverse geometry is almost rapidity independent
\cite{Dumitru:2008wn}.

\bigskip

\noindent\fbox{
\parbox{\textwidth}{ {\bf Exercise 8}: Make three different plots of
  $\cos(2 \Delta \phi)$, $\cos(2 \Delta \phi)+0.25\cos(3 \Delta \phi)$
  and $\cos(2 \Delta \phi)+0.55\cos(3 \Delta \phi)$, in the range
  $\Delta \phi = -3\pi/2$ to $\pi/2$. Notice how the heights and
  widths of the two peaks change as one increases the admixture of
  $v_3$. Notice also the appearance of a ``shoulder'' in the away-side
  peak. Compare with Fig. 5 of \cite{CMS:2013bza} which is similar to
  our Fig. \ref{ridge1}(a), but for ultracentral (0-0.2 \%) collisions, and
  clearly shows the shoulder formation on the away side.}}

\bigskip

Interestingly, the ridge is also observed in high-multiplicity pp
collisions (Fig. \ref{ridge2}). This begs the question: is quark-gluon
plasma formed or is hydrodynamic flow developed also in
high-multiplicity pp (and p-nucleus) collisions? This issue is still
being debated (see Sec. 4).


\subsection{Flow decorrelation and factorization breaking}

Consider the two-particle Fourier coefficient 
$V_{n\Delta}=\mean{\exp(in\Delta\phi)}$ 
occurring in Eq. (\ref{dNdDphi}). Making the
same assumptions as in the derivation of Eq. (\ref{mpc3}), one gets
$V_{n\Delta}=v_n^2$.  However, as stated in Sec. 2.2, the event-plane
angle $\Psi_n$ may depend on the transverse momentum $(p_T)$ and
pseudorapidity $(\eta)$, and hence may not be the same in the two bins $a$
and $b$ from which the two particles are taken.
In this case, $V_{n\Delta}$ in a single event
is given by
\begin{eqnarray}
\left.\begin{aligned}
V_{n\Delta}(a,b)=\mean{e^{i n (\phi^a-\phi^b)}}=\mean{e^{i n \phi^a}}
\mean{e^{-i n \phi^b}}=V_{n,a}V_{n,b}^*=v_{n,a}v_{n,b}\,
e^{i n (\Psi_n^a-\Psi_n^b)},\nonumber
\end{aligned}\right.\end{eqnarray}
where $\mean{\cdots}$ denotes the average over one event
and the second equality is obtained by assuming that there are no
nonflow correlations. 
Factorization (of two-particle correlations) refers to
$V_{n\Delta}(a,b)$ getting factorized into a product of
single-particle harmonics, $v_{n,a}$ and $v_{n,b}$:
$V_{n\Delta}(a,b)=v_{n,a} \times v_{n,b}$. Factorization breaks down
($V_{n\Delta}(a,b) \neq v_{n,a} \times v_{n,b}$) if $\Psi_n$ is not
the same in the two bins and/or if there are nonflow correlations. The
latter effect can be minimized by introducing a large-enough
pseudorapidity gap between the two bins.

Experimentally, one further averages over many
events to write
\begin{eqnarray}
\left.\begin{aligned}
V_{n\Delta}(a,b)=\mean{V_{n,a}V_{n,b}^*}=\mean{v_{n,a}v_{n,b}\,
e^{i n (\Psi_n^a-\Psi_n^b)}},
\end{aligned}\right.\end{eqnarray}
where $\mean{\cdots}$ denotes the average over many events.
Factorization ratio $r_n(p_T^a,p_T^b)$ is defined as
\cite{Gardim:2012im}
\begin{equation}
\label{fact2}
r_n(p_T^a,p_T^b) \equiv \frac{ V_{n\Delta}(p_T^a,p_T^b)}
{\sqrt{ V_{n\Delta}(p_T^a,p_T^a)V_{n\Delta}(p_T^b,p_T^b)}}=
\frac{\mean{v_n(p_T^a) v_n(p_T^b)\cos n(\Psi_n(p_T^a)-\Psi_n(p_T^b))}}
{\sqrt{\mean{v_n^2(p_T^a)}\mean{v_n^2(p_T^b)}}}.
\end{equation}
It measures the flow decorrelation between momentum bins, i.e.,
$p_T$-dependent event-plane angle fluctuations. If $V_{n\Delta}$
factorizes, $r_n$ is obviously unity, otherwise it is less than unity
by the Cauchy–-Schwarz inequality. In the most central events
initial-state fluctuations play a dominant role, and so $r_n$ deviates
from unity significantly. $r_n(p_T^a,p_T^b)$ is a sensitive
discriminator of the models of the initial-state geometry in the
transverse plane. Specifically, it could probe the granularity or
spatial extent of the density fluctuations in the transverse plane in
the initial state.

Assuming $V_{n\Delta}(p_T^a,p_T^b)$ factorizes into the product of
single-particle anisotropies, $p_T$ dependence of the latter can be
determined experimentally from
\begin{equation}
v_n(p_T^a) = \frac{V_{n\Delta}(p_T^a,p_T^b)}
{\sqrt{V_{n\Delta}(p_T^b,p_T^b)}},
\end{equation}
where the reference momentum $p_T^b$ is chosen not too large to
minimize correlations from jets at higher $p_T$.

The longitudinal decorrelation or $\eta$-dependent factorization
breakdown cannot be studied by simply replacing $p_T^a$ and $p_T^b$ in
Eq. (\ref{fact2}) by $\eta_a$ and $\eta_b$, respectively, because the
denominator would be badly contaminated by short-range nonflow
correlations. Instead, one defines the following observable:
\begin{eqnarray}\label{decor}\left.\begin{aligned}
r_n(\eta) & \equiv \frac{\mean{V_n(-\eta)V_n^*(\eta_{ref})}}
{\mean{V_n(+\eta)V_n^*(\eta_{\rm ref})}}\\ &=
\frac{\mean{v_n(-\eta)v_n(\eta_{\rm ref})\cos
    n(\Psi_n(-\eta)-\Psi_n(\eta_{\rm ref}))}}
     {\mean{v_n(+\eta)v_n(\eta_{\rm ref})\cos
         n(\Psi_n(+\eta)-\Psi_n(\eta_{\rm
           ref}))}},
\end{aligned}\right.\end{eqnarray}
with a large gap between $\eta$ and $\eta_{\rm ref}$.
Experimentally, it is found that $v_n$ and $\Psi_n$ fluctuate along
the longitudinal direction, even in a given event
\cite{Khachatryan:2015oea,Aaboud:2017tql}.  $r_n(\eta)$ is unity at
$\eta=0$ and decreases as $\eta$ increases
\cite{Aad:2020gfz}. Hydrodynamic model calculations show that
$r_n(\eta)$ is driven mostly by the longitudinal structure of the
primordial state. The study of $r_n(\eta)$ puts constraints on the
complete three-dimensional modelling of the initial state used in 
3+1D models of the system evolution. 
A significant breakdown of factorization in $\eta$ is also observed 
in pPb collisions \cite{Khachatryan:2015oea}.

\medskip
To conclude, experimentally, factorization breakdown or flow
decorrelation has been observed as a function of both $p_T$ and $\eta$
\cite{Khachatryan:2015oea,CMS:2013bza,Aaboud:2017tql}.
For $v_2$ in central PbPb collisions at LHC the effect can be as large as 20\%.


\section{Observables with Mixed Harmonics}

Flow observables discussed so far referred to a single harmonic
$n$. We now discuss flow observables with mixed harmonics. These
provide new handles on the initial state as well as the properties of
the medium. Specifically, they can discriminate among the various
initial-state models as well as among different parametrizations of
the temperature dependence of the specific shear ($\eta/s$) and bulk
viscosity ($\zeta/s$); $s$ is the entropy density.

\medskip

In Sec. 2.3 we discussed multiparticle correlations involving only a single
harmonic ($n$). The most general $k$-particle correlation involving
harmonics $n_1,n_2,\cdots,n_k$ can be written as \cite{Bhalerao:2011yg}
\begin{equation}
\label{mpc7}
v\{n_1,n_2,\cdots,n_k\}=\langle\langle e^{i(n_1\phi_1+\cdots +n_k\phi_k)}
\rangle\rangle,
\end{equation}
where $n_1,n_2,\cdots,n_k$ are integers satisfying
$n_1+n_2+\cdots+n_k=0$, so that this correlation observable is
invariant under rotation in the azimuthal plane. The notation
$\langle\langle \cdots \rangle\rangle$ is as in Eq. (\ref{mpc1}).
If there are no nonflow correlations, this becomes
\begin{equation}
\label{mpc8}
v\{n_1,n_2,\cdots,n_k\}=\langle\langle e^{in_1\phi_1}\rangle
 \cdots \langle e^{in_k\phi_k}
\rangle\rangle = \langle v_{n_1} \cdots v_{n_k} 
e^{i(n_1\Psi_{n_1}+\cdots +n_k\Psi_{n_k})} \rangle,
\end{equation}
where in the final expression the average is only over events. Using
this, a number of new flow observables were constructed to study mixed
correlations between $v_1,\,v_2$ and $v_3$ 
\cite{Bhalerao:2011yg,Bhalerao:2011ry}.


\subsection{Symmetric cumulants}

Symmetric cumulants $SC(m,n)$ and $NSC(m,n)$ measure the correlations
between event-by-event fluctuations of
the {\it magnitudes} of the complex flow vectors $V_m$ and $V_n$ ($m \ne n$):
\begin{eqnarray}\label{symcum}\left.\begin{aligned}
SC(m,n)&\equiv \mean{v_m^2 v_n^2}-\mean{v_m^2}\mean{v_n^2},\\
NSC(m,n)&\equiv \frac{\mean{v_m^2 v_n^2}-\mean{v_m^2}\mean{v_n^2}}{\mean{v_m^2}
\mean{v_n^2}},
\end{aligned}\right.\end{eqnarray}
which obviously vanish if $v_m^2$ and $v_n^2$ are uncorrelated or if there
are no flow fluctuations.
$NSC(m,n)$ can be looked upon as the mixed kurtosis of $v_m$ and $v_n$
(see Eq. (\ref{A3})). Hydrodynamic calculations show that $v_n$
scales approximately linearly with $\varepsilon_n$, for $n=2,3$. An
advantage in defining the ratio as in Eq. (\ref{symcum}) 
is that the constants of proportionality between $v_n$ and
$\varepsilon_n$ drop out, thereby directly connecting experimental
observables with the properties of the initial state. Symmetric
cumulant is a four-particle observable.

\bigskip

\noindent\fbox{\parbox{\textwidth}{ {\bf Exercise 9}: Equation
    (\ref{mpc3}) together with the discussion following it, explain
    how $\mean{v_n^2}$ and $\mean{v_n^4}$ can be measured. Explain how
    you would measure $\mean{v_m^2 v_n^2}$. Note also that, as in
    Eq. (\ref{mpc3}), any dependence on the symmetry plane is
    eliminated by construction.  }}

\bigskip

\begin{figure}
\begin{center}
\resizebox{0.6\columnwidth}{!}{\includegraphics{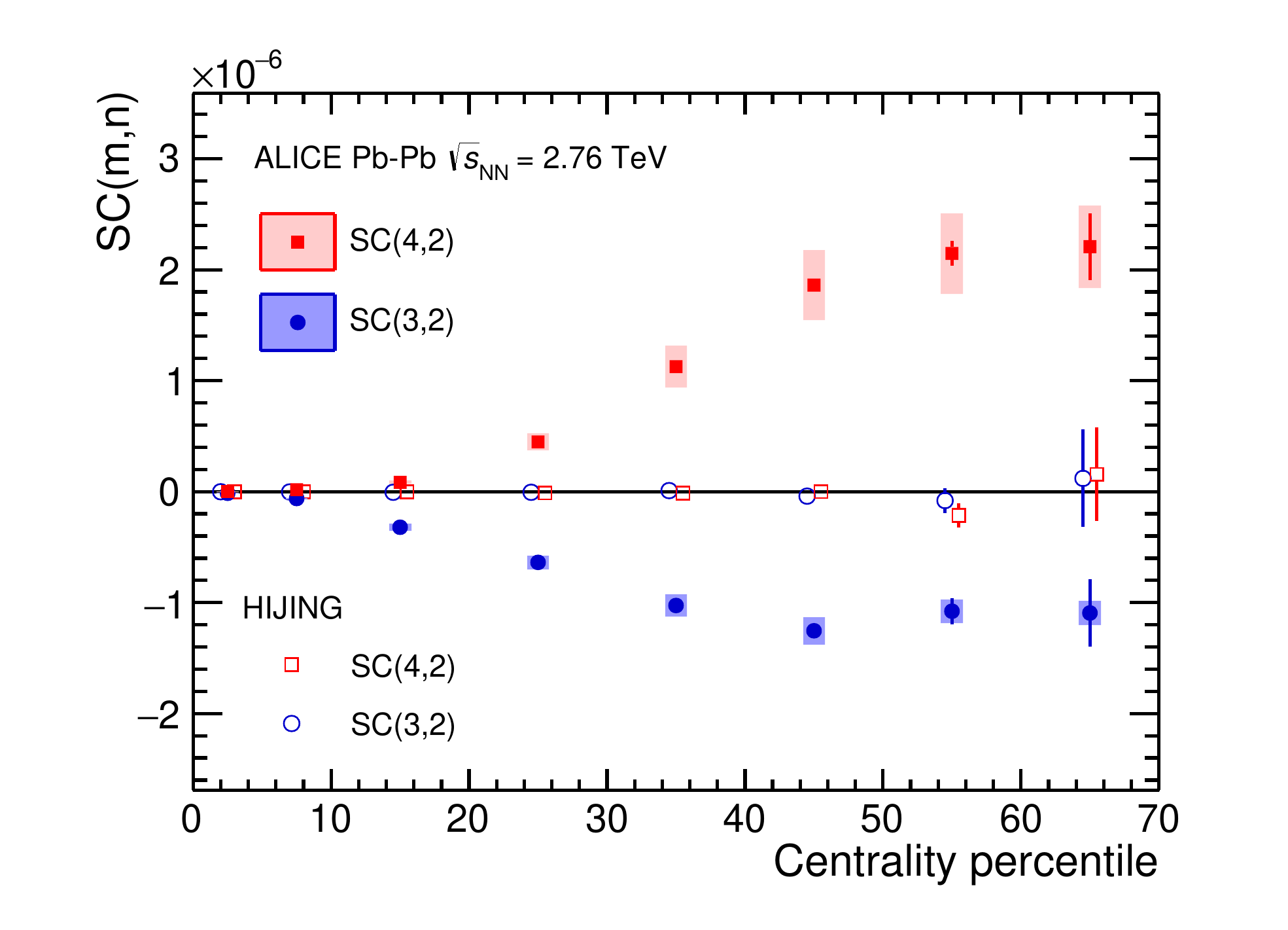}}
\caption{(colour online) Centrality dependence of the symmetric
  cumulants $SC(4,2)$ (red filled squares) and $SC(3,2)$ (blue filled
  circles). Boxes represent systematical errors. HIJING model results
  are shown with hollow markers. Figure from \cite{ALICE:2016kpq}.}
\label{scmn}
\end{center}
\end{figure}

Figure \ref{scmn} shows the ALICE results for $SC(4,2)$ and $SC(3,2)$
as a function of centrality. $SC(4,2)$ is found to be positive and
$SC(3,2)$ negative. This indicates that if in an event $v_2$ is larger
than $\mean{v_2}$, then there is an enhanced probability of finding
$v_4$ larger than $\mean{v_4}$. Similarly, if in an event $v_2$ is
larger than $\mean{v_2}$, then there is an enhanced probability of
finding $v_3$ {\it smaller} than $\mean{v_3}$. In other words, $v_2$
and $v_4$ are correlated, whereas $v_2$ and $v_3$ are anticorrelated,
for all centralities. The detailed behaviour, however, depends on whether
one is in the fluctuation-dominated (most central) regime or 
the geometry-dominated (midcentral) regime.
Figure \ref{scmn} also shows the HIJING model results which are
consistent with zero. Now, this model includes nonflow azimuthal
correlations due to jet production, but has no physics of
collectivity. Evidently, the nonzero ALICE results cannot be explained
by nonflow effects, showing that the symmetric cumulants $SC(m,n)$ are robust
against systematic biases originating from nonflow effects. More
recently, ALICE have presented data on $SC(4,3)$, $SC(5,2)$ and
$SC(5,3)$ \cite{Acharya:2017gsw}.
Interestingly, negative $SC(3,2)$ has been observed also in the
highest-multiplicity pp and pPb events
\cite{Sirunyan:2017uyl,Acharya:2019vdf}.

Recently, the idea of symmetric cumulants was generalized and a
new set of observables, which quantify the correlations of flow
fluctuations involving {\it more than two} flow amplitudes, was
introduced \cite{Mordasini:2019hut}.

\subsection{Event-plane correlators}

Event-plane correlators or symmetry-plane correlators
measure the correlations between
event-by-event fluctuations of the {\it phases} of the complex flow
vectors $V_m$ and $V_n$ ($m \ne n$). They represent higher-order
correlations, involving at least three particles (see below). One can
construct two-, three-, or higher-plane correlators. They thus
bring in a large number of new observables that provide new, detailed
insight into the hydrodynamic response and the initial state.

Suppose we want to construct an observable that measures correlations
between the event planes $\Psi_m$ and $\Psi_n$. Recall first that
$\Psi_l$ can be determined only modulo $2\pi/l$
(Eq. (\ref{Psin1})). Hence the observable should be invariant under
$\Psi_l \rightarrow \Psi_l+2\pi/l$. Moreover, it should also be
invariant under a global rotation through an arbitrary angle. An
observable that satisfies both these requirements is $\mean{\cos
  k(\Psi_m-\Psi_n)}$, where $k$ is a common multiple of $m$ and $n$.
Consider the simplest example\footnote{Recall the discussion in the
last paragraph of Sec. 2.3.} of $\mean{\cos 4(\Psi_4-\Psi_2)}$.
In order to measure this correlator, we consider 
\begin{eqnarray}\label{obs1}\left.\begin{aligned}
\frac{{\rm Re} \mean{V_4V_2^{\ast2}}}{\sqrt{\mean{|V_4|^2}\mean{|V_2|^4}}} 
&= \frac{\mean{v_4 v_2^2 \cos 4(\Psi_4-\Psi_2)}}{\sqrt{\mean{v_4^2}\mean{v_2^4}}}
\approx \mean{ \cos4(\Psi_4-\Psi_2)},
\end{aligned}\right.\end{eqnarray}
where $V_n$ is the experimentally measured flow vector, often denoted
by $Q_n$ or $q_n$ (see Sec. 2.2.). It is clear that the left-hand side of the above equation
involves averaging $\exp (2i\phi_j+2i\phi_k-4i\phi_l)$ over triplets
of particles in an event and then over many events, and thus
necessarily involves three-particle correlations.
In practice, the three particles are taken from two or more different
pseudorapidity bins to minimize the short-range nonflow correlations.
Observables for other event-plane correlators can be constructed
similarly \cite{Bhalerao:2013ina}.\footnote{ATLAS collaboration denotes the
  event plane by $\Phi_n$, which is different from our convention here. In
  Ref. \cite{Bhalerao:2013ina}, we have used the ATLAS convention.}

\bigskip

\noindent\fbox{\parbox{\textwidth}{ {\bf Exercise 10}: Table I in
    Ref. \cite{Bhalerao:2013ina} lists two-, three- and four-plane
    correlators, along with the total number of particles that are
    involved in each case. For example, the two-event-plane correlator
    $\mean{\cos 12(\Psi_2-\Psi_4)}$, the three-event-plane correlator
    $\mean{\cos(10\Psi_2-6\Psi_3-4\Psi_4)}$ and the four-event-plane
    correlator $\mean{\cos(6\Psi_2+3\Psi_3-4\Psi_4-5\Psi_5)}$, involve 9, 8
    and 6 particles, respectively. Understand this Table.}}

\bigskip

\begin{figure}
\resizebox{0.99\columnwidth}{!}{\includegraphics{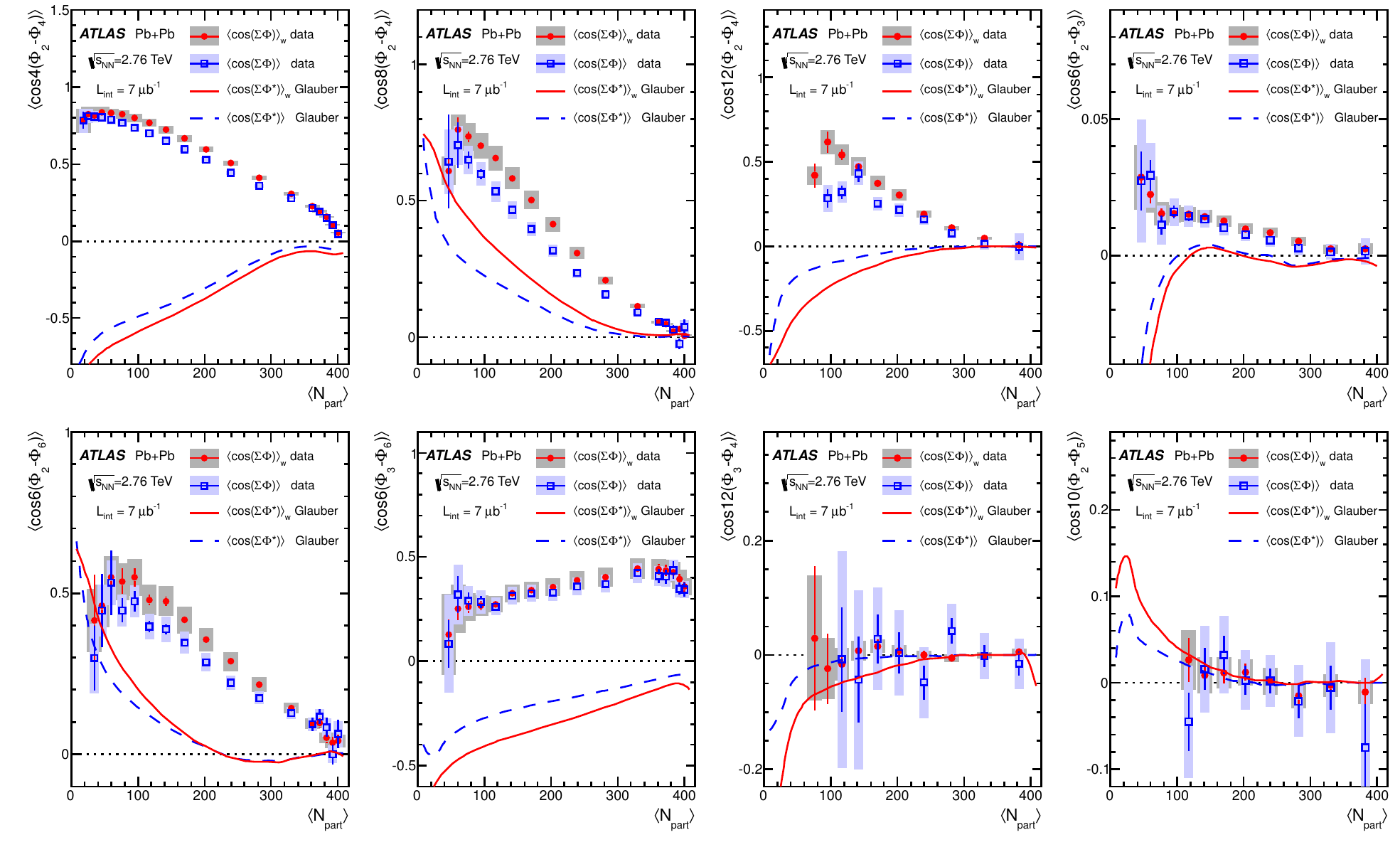}}
\caption{(colour online) ATLAS data \cite{Aad:2014fla} on
  two-event-plane correlators as a function of centrality. Red solid
  symbols and blue open symbols represent two different analysis
  methods. Red solid and blue dashed curves represent Glauber-model
  predictions for two-participant-plane correlators using two
  different methods.  Here $\Phi_n$ denotes the event-plane and
  $\Phi_n^*$ the participant plane, which differs from our
  convention.}
\label{epc}
\end{figure}

Figure \ref{epc} shows the ATLAS data on eight two-plane correlators
as a function of centrality for Pb-Pb collisions at $\sqrt{s_{\rm
    NN}}=2.76$ TeV \cite{Aad:2014fla}. Given a model for the initial
state, say the Glauber model, one can calculate participant-plane
correlators as a function of centrality; these are also shown in
Fig. \ref{epc}. 
Here the participant-plane correlators and event-plane correlators are
being compared with each other to illustrate the role played by the
hydrodynamic evolution in converting the initial-state spatial
correlations between different harmonics into the corresponding
final-state momentum-space correlations. For the most central events,
where fluctuations and not geometry is the dominant feature, the
two-plane (Fig. \ref{epc}) as well as three-plane \cite{Aad:2014fla}
correlators vanish (with the sole exception of $\mean{\cos
  6(\Phi_3-\Phi_6)}$). For off-central collisions, geometry takes
over, and the correlators increase in magnitude. This is similar to
the centrality dependence of the symmetric cumulants
(Fig. \ref{scmn}). For off-central collisions, the participant-plane
correlators and event-plane correlators generally differ
significantly, especially for higher harmonics, pointing to the
misalignment between $\Psi_n$ and  $\Phi_n$, owing to the
contributions of the nonlinear flow modes, to be discussed in the next
subsection.

ALICE collaboration has measured a few two- and
three-plane correlators; they denote them generically by $\rho_{n,mk}$
\cite{Acharya:2020taj,Acharya:2017zfg}.
A recent paper has pointed out problems with the current methods of
measuring symmetry-plane correlations and has proposed a new estimator
as an improvement \cite{Bilandzic:2020csw}.

\subsection{Nonlinear flow modes and mode coupling or mixing}

Hydrodynamic calculations show that $V_n$ scales approximately
linearly with the initial eccentricity $\E_n$, for
$n=2,3$. For higher harmonics, however, $V_n$ receives contributions
from eccentricities in the lower harmonics as well \cite{Teaney:2012ke}.
For complex flow
vectors $V_n \, (n \ge 4)$ one can write the vector sum of the linear and
nonlinear response terms as \cite{Acharya:2020taj,Acharya:2017zfg,Qian:2016fpi}.
\begin{equation}
\label{nonlin}
\begin{split}
        V_4&=V_{4\mathrm{L}}+V_{4\mathrm{NL}}=V_{4\mathrm{L}}+\chi_{4,22} V_2^2,\\
        V_5&=V_{5\mathrm{L}}+V_{5\mathrm{NL}}=V_{5\mathrm{L}}+\chi_{5,23} V_2 V_3,\\
        V_6&=V_{6\mathrm{L}}+V_{6\mathrm{NL}}=V_{6\mathrm{L}}+\chi_{6,222} V_2^3+\chi_{6,33} V_3^2
+\chi_{6,24} V_2 V_{4L},\\
        V_7&=V_{7\mathrm{L}}+V_{7\mathrm{NL}}=V_{7\mathrm{L}}+\chi_{7,223} V_2^2 V_3+\chi_{7,34} V_3 V_{4L}
+\chi_{7,25} V_2 V_{5L},\\
        V_8&=V_{8\mathrm{L}}+V_{8\mathrm{NL}}=V_{8\mathrm{L}}+\chi_{8,2222} V_2^4+\chi_{8,233} V_2 V_3^2
+\mathcal{E}(V_{4L},V_{5L},V_{6L}),
\end{split}
\end{equation}
where the first term on the right in each equation is called the
linear flow mode and the remaining terms constitute the nonlinear flow
modes. The (real) coefficients $\chi$ are called the nonlinear flow mode
coefficients, and $\mathcal{E}(\cdots)$ in the last equation denotes
the remaining terms in $V_8$.

Now, suppose the linear terms $V_{4L},\,V_{5L},\,V_{6L}$, etc. are
{\it defined} to be linearly proportional to the corresponding
cumulant-based eccentricities $\E_4,\, \E_5,\, \E_6$, etc. defined in
Eq. (\ref{eccn3}), rather than the moment-based eccentricities defined
in Eq. (\ref{eccn1}). For example, $V_{4L}$ will be driven not by
$\varepsilon_4 \exp(i4\Phi_4)$, but by
\[
- \frac{\mean{z^4}- 3 \mean{z^2}^2 }{\mean{r^4}}
=\varepsilon_4 e^{i4\Phi_4}+\frac{3\mean{r^2}^2}{\mean{r^4}}
\varepsilon_2^2 e^{i4\Phi_2}.
\]
With the above definition, the linear and nonlinear parts in
Eqs. (\ref{nonlin}) are not necessarily uncorrelated.  On the other
hand, if the linear term $V_{nL}$ in each of the Eqs. (\ref{nonlin})
is {\it defined} as the part of $V_n$ that is uncorrelated with the
nonlinear part $V_{nNL}$, then the linear term may or may not be
proportional to the corresponding cumulant-based eccentricity.
Clearly, more work needs to be done here for a better
understanding. In what follows we describe a few recent works, so
that the reader can follow the literature.

Let us first assume that the coefficients $\chi$ are the same
for all events in a centrality class.\footnote{Note also that
  $\mean{V_{nL}}=0$, because $V_{nL}$, like $V_n$, is expected to
  carry a random phase factor depending on the reaction-plane angle
  $\Phi_{\rm RP}$.}
Now if the linear and nonlinear parts in Eq. (\ref{nonlin}) are
uncorrelated, then it is easy to show that \cite{Bhalerao:2014xra}
\begin{eqnarray}\label{}\left.\begin{aligned}
\frac{\mean{V_4 V_2^{\ast 2}v_2^2}}{\mean{V_4 V_2^{\ast 2}}\mean{v_2^2}}
&= \frac{\mean{v_2^6}}{\mean{v_2^4}\mean{v_2^2}}, \\
\frac{\mean{V_5 V_2^\ast V_3^\ast v_2^2}}{\mean{V_5 V_2^\ast V_3^\ast}\mean{v_2^2}}
&= \frac{\mean{v_2^4 v_3^2}}{\mean{v_2^2 v_3^2}\mean{v_2^2}}.
\end{aligned}\right.\end{eqnarray}
The left- and right-hand sides of these and other similar equations
are found to agree with each other to a good approximation in AMPT, as
well as in perfect and imperfect hydrodynamic calculations
\cite{Qian:2016fpi,Bhalerao:2014xra}, justifying the hypothesis that
the linear and nonlinear contributions are uncorrelated.

To isolate the linear and nonlinear terms in Eq. (\ref{nonlin}), it is
usually assumed that they are uncorrelated \cite{Yan:2015jma}. Consider, for example,
the first of the five equations in (\ref{nonlin}), take its complex
conjugate, multiply the two, average over many events in a centrality
class, to get
\[
\mean{v^2_{4\mathrm{L}}} = \mean{v^2_4} - \chi^2_{4,22} \mean{v^4_2}.
\]
This, however, cannot be used to determine the linear part because
$\chi_{4,22}$ is an unknown. Consider, however, the observable
\begin{equation}\label{obs2}
\frac{[\mathrm{Re}\mean{V_4 V_2^{\ast 2}}]^2}{\mean{v_2^4}}=
\chi^2_{4,22} \mean{v^4_2}.
\end{equation}
It allows us to determine the linear part as well as the coefficient
$\chi^2_{4,22}$ \cite{Yan:2015jma}:
\begin{equation}\label{NLcoeff}
\mean{v^2_{4\mathrm{L}}} = \mean{v^2_4} 
- \frac{[\mathrm{Re}\mean{V_4 V_2^{\ast 2}}]^2}{\mean{v_2^4}}
~~~~\mathrm{and}~~~ \chi_{4,22}=
\frac{\mathrm{Re}\mean{V_4 V_2^{\ast 2}}}{\mean{v_2^4}}.
\end{equation}
Similarly, it can be shown that
\begin{equation}\label{}
\mean{v^2_{5\mathrm{L}}} = \mean{v^2_5} 
- \frac{[\mathrm{Re}\mean{V_5 V_2^{\ast}V_3^{\ast}}]^2}{\mean{v_2^2 v_3^2}}
~~~~\mathrm{and}~~~ \chi_{5,23}=
\frac{\mathrm{Re}\mean{V_5 V_2^{\ast}V_3^{\ast}}}{\mean{v_2^2 v_3^2}}.
\end{equation}
Other equations in (\ref{nonlin}) can be treated similarly; see
\cite{Acharya:2020taj,Qian:2016fpi,Yan:2015jma,Qian:2017ier} for details.

The nonlinear flow mode coefficients ($\chi$) are expected to be
independent of the initial density profile in a given centrality
class. Centrality dependence of these coefficients has been presented
in Refs. \cite{Acharya:2017zfg} and \cite{Acharya:2020taj}, for PbPb
collisions at 2.76 and 5.02 TeV, respectively.


\subsubsection{Connections between seemingly unrelated observables}

We now point out an interesting connection between the two seemingly
unrelated observables, namely, event-plane correlators (Sec. 3.2) and
the nonlinear response (Sec. 3.3): This is evident in the similarity
between the observables defined in Eqs. (\ref{obs1}) and (\ref{obs2}),
which allows us to write $\mean{\cos4(\Psi_4-\Psi_2)}$ in terms of
$\chi_{4,22}$. These equations illustrate how the event-plane
correlations can be understood in the framework of the nonlinear
response. Correlations between the {\it magnitudes} of the flow
vectors of different harmonics also exhibit distinctive signs of the
nonlinear mode couplings \cite{Aad:2015lwa}.

Consider Eqs. (\ref{nonlin}). One can write similar nonlinear
equations connecting the initial eccentricity vectors $\E_n ~(n \ge 4)$ with
$\E_m$ of lower harmonics ($m<n$), with coefficients $\chi$ replaced
by, say, $\tilde{\chi}$. An interesting question is how the nonlinear
mode coupling coefficients in the two cases, $\chi$ and
$\tilde{\chi}$, are related. This has been investigated in
\cite{Qian:2016fpi}. Just as the nonlinear-response terms in the
higher-harmonic flow vectors give rise to event-plane correlations
in the final state, the nonlinear terms in the higher-harmonic
eccentricity vectors cause participant-plane correlations in the
initial state. Relationships between $\chi$ and $\tilde{\chi}$ are
reflected in the relationships between event-plane and
participant-plane correlations \cite{Qian:2016fpi}.


\section{Collectivity}

Collectivity refers to a large number of particles acting in
unison. Multiparticle cumulants discussed in Sec. 2 probe just this
aspect of the data (Figs. \ref{v2m}, \ref{ng3}). Symmetric cumulants
$SC(m,n)$ are sensitive to four-particle correlations (Exercise
9). Event-plane correlators also probe multiparticle correlations
(Exercise 10). Collectivity is also evident in the ridge phenomenon ---
the azimuthally collimated, near-side, long-range rapidity
correlations between two hadrons (Figs. \ref{ridge1} and
\ref{ridge2}). All the above signatures of collectivity have been seen
very clearly in the large systems formed in relativistic
heavy-ion collisions.

Traditionally, small systems such as those formed in pp or p-nucleus
collisions were considered as a benchmark or reference for studying
large systems formed in nucleus-nucleus collisions, the assumption
being that the quark-gluon plasma was unlikely to be produced in small
systems. However, with the small systems displaying qualitatively
similar behaviour as the large systems, this assumption becomes
questionable.

\begin{figure}
\begin{center}
\resizebox{0.99\columnwidth}{!}{\includegraphics{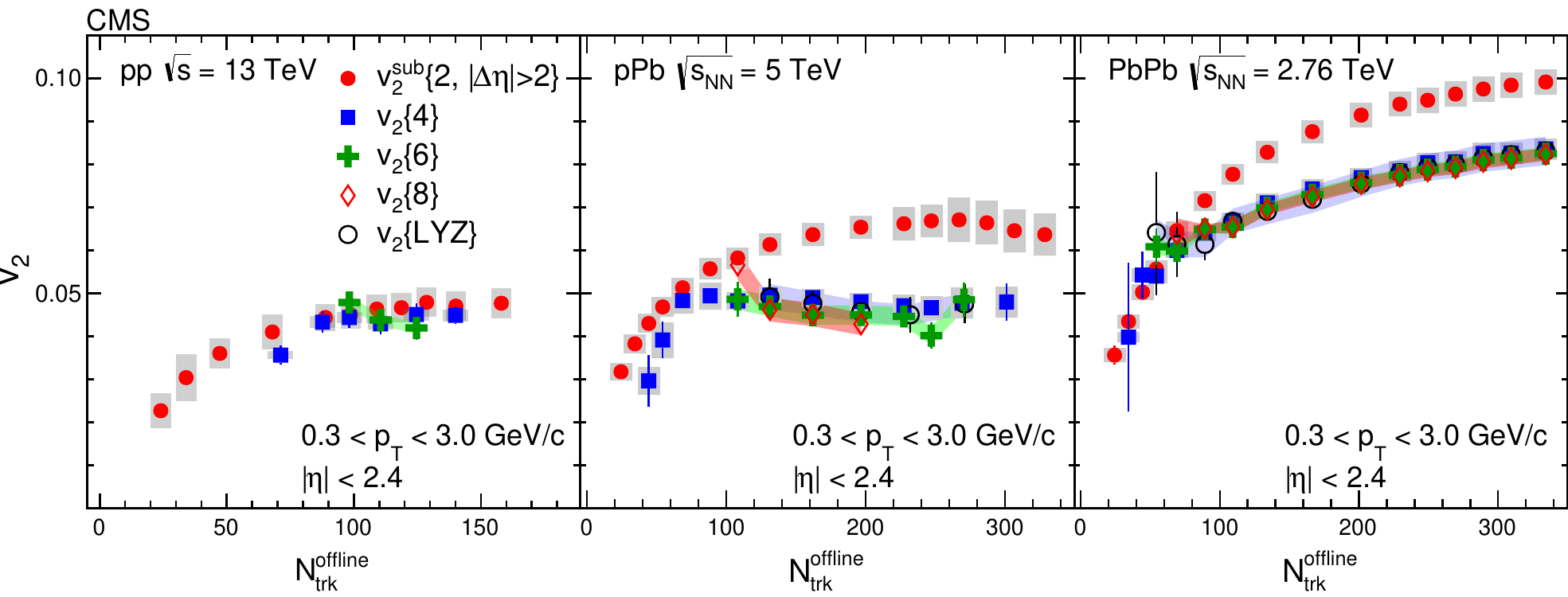}}
\caption{(colour online) Observation of multiparticle cumulants in pp,
  pPb and PbPb collisions. Points labelled $v_2\{\rm LYZ\}$ refer to the analysis
method based on the Lee-Yang zeroes 
\cite{Bhalerao:2003yq,Bhalerao:2003xf}.
The error bars correspond to the statistical
  uncertainties, while the shaded areas denote the systematic
  uncertainties. Figure from \cite{Khachatryan:2016txc}.}
\label{pppPb}
\end{center}
\end{figure}

We saw in Sec. II that ridges have been observed even in small systems
created in high-multiplicity pp and p-nucleus collisions. Ridge was a crucial piece of
evidence for the strong collective behaviour comparable to a
fluid in heavy-ion collisions. Some other observables or features of the 
data that were
thought to be consistent with the formation of quark-gluon plasma and
were originally observed in large systems, have now been observed in
small systems as well. Among them are: sizable azimuthal anisotropy
--- not just $v_2$, but also higher harmonics --- in the
highest-multiplicity events \cite{Khachatryan:2016txc,Aaboud:2016yar},
mass ordering of $v_2(p_T)$ of identified particles
\cite{Pacik:2018gix}, multiparticle cumulants (Fig. \ref{pppPb})
\cite{Khachatryan:2016txc},
system size and shape dependence of the functions $v_2(p_T)$ and
$v_3(p_T)$ \cite{PHENIX:2018lia}, enhanced strangeness production
\cite{ALICE:2017jyt},
negative $SC(3,2)$ in the
highest-multiplicity pp and pPb events
\cite{Sirunyan:2017uyl,Acharya:2019vdf}, etc. Azimuthal anisotropy suggests
geometry-driven pressure gradients, mass ordering signifies a common
velocity field of the expanding medium, ridge and multiparticle cumulants point
toward a collective behaviour, strangeness enhancement indicates a
deconfined medium --- all leading to the proposition that perhaps
``One fluid rules them all ...  pp, pPb, PbPb'' \cite{Weller:2017tsr}.
However, this claim has been contested (``One fluid might not rule
them all'') on the grounds that the 2+1D hydrodynamic simulations were
unable to describe the multiparticle single and mixed harmonics
cumulants \cite{Zhou:2020pai}.
Moreover, some other observables such as jet quenching, high-$p_T$
hadron suppression, bottomonium suppression, etc., seen in large
systems have remain elusive in small-system collisions
\cite{Acharya:2018qsh}. So the question remains: what is the
smallest size of a drop of liquid QGP? Or, can one dial the system
size and switch off/on the QGP?


\subsection{Origin of collectivity}

\begin{figure}
\begin{center}
\resizebox{0.99\columnwidth}{!}{\includegraphics{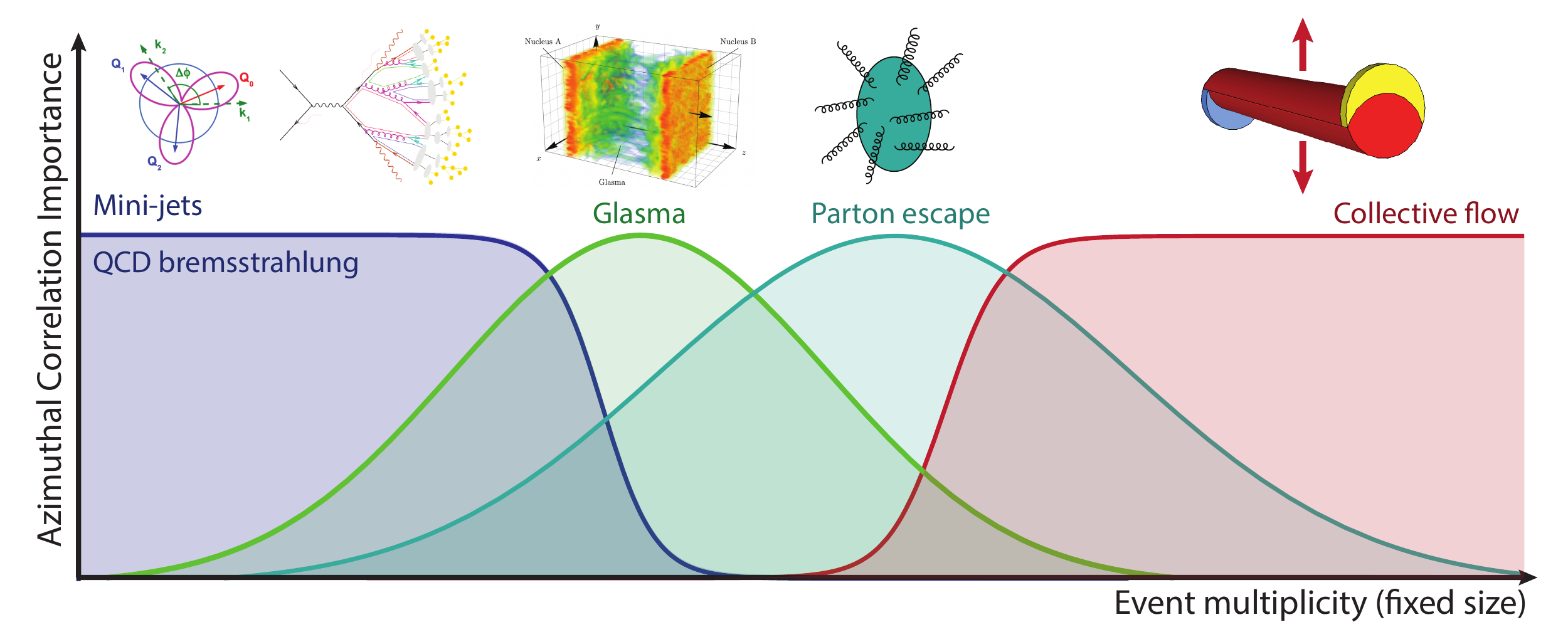}}
\caption{(colour online) A schematic diagram depicting various sources
  of azimuthal anisotropy as a function of event
  multiplicity. Vertical scale is arbitrary. Figure from
  \cite{Strickland:2018exs}.}
\label{strickland}
\end{center}
\end{figure}

Azimuthal anisotropy has long been considered as a signature of
transverse collective flow \cite{Ollitrault:1992bk}. It is generally
agreed that the azimuthal anisotropy (for $p_T \lesssim 2$ GeV) in
relativistic heavy-ion collisions is of hydrodynamic origin. However,
at lower event multiplicities, the anisotropy can have a variety of
other (nonhydrodynamic) sources (Fig. \ref{strickland}); for an
excellent brief review see \cite{Strickland:2018exs}. We shall outline
only three of these competing theoretical scenarios below.


\subsubsection{Hydrodynamics}

\begin{figure}
\begin{center}
\resizebox{0.45\columnwidth}{!}{\includegraphics{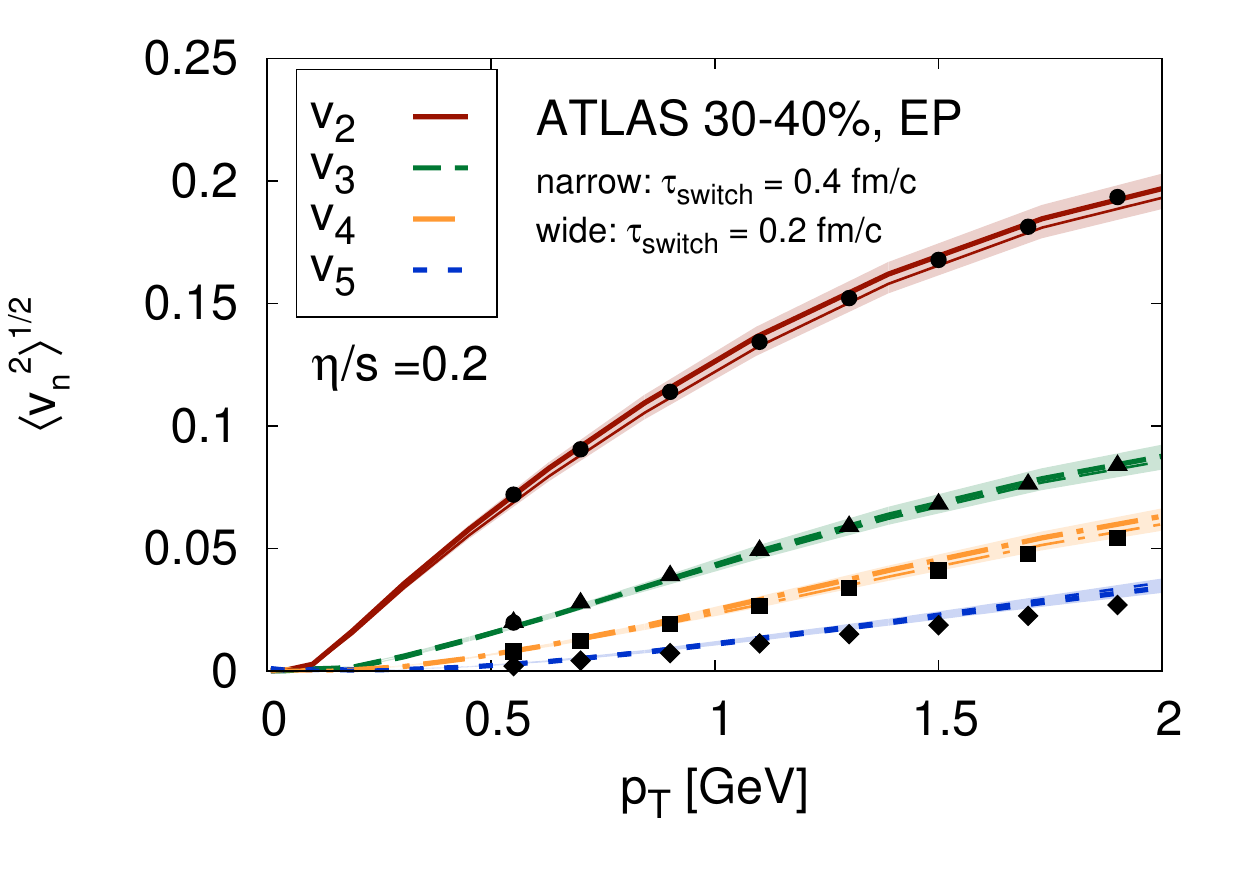}}
\caption{(colour online) Root-mean-square flow (Eq. \ref{mpc5}) versus
  transverse momentum. Hydrodynamic calculations with two different
  switching times are compared to the ATLAS data \cite{ATLAS:2012at}
  obtained using the event-plane (EP) method. Bands indicate
  statistical errors. Experimental error bars are smaller than the
  size of the points. Figure from \cite{Gale:2012rq}.}
\label{Gale}
\end{center}
\end{figure}

Hydrodynamic response converts the initial spatial anisotropy of the
nuclear overlap region into the momentum-space anisotropy of the
final-state particles. Event-by-event relativistic second-order
dissipative hydrodynamics has been successful in explaining most of
the flow-related data. As an example of the success of the
hydrodynamic paradigm, see Fig. \ref{Gale}. Note also the ordering:
$v_2>v_3>v_4>v_5$.\footnote{A recent experiment \cite{Acharya:2020taj}
  has shown that the ordering persists up to $n=7$, with some
  enhancement for $n=8,9$.} It arises because, for the 30-40\%
centrality results shown in Fig. \ref{Gale}, the shape of the initial
overlap area is predominantly elliptic. The higher harmonics are
driven mostly by the eccentricity fluctuations. As they probe finer
details of the initial density distribution, they are successively
smaller. The ordering is seen even in ideal hydrodynamic calculations;
see, e.g., \cite{Gardim:2012yp}. Viscous hydrodynamic calculations
enhance the ordering because the higher harmonics are found to be more
sensitive to the viscous attenuation. For the ultracentral collisions,
the flow is driven mainly by the fluctuations rather than by the
geometry, and so $v_2$ and $v_3$ are found to be similar in
magnitude. Higher harmonics receive contributions from nonlinear flow
modes (Sec. 3.3) and so their interpretation is not very
straightforward.

Fluid dynamics (or loosely speaking, hydrodynamics) is an effective
(macroscopic) theory that describes the slow, long-wavelength motion
of a fluid close to local equilibrium. However, in the context of
relativistic heavy-ion collisions, hydrodynamics is found to work 
very well even
when the pressure in the system is far from isotropization ($p_L \ne
p_T$). Secondly, applicability of hydrodynamics requires a clear
separation between the microscopic and macroscopic length or time
scales. Traditionally, the theory is formulated as a systematic
expansion in gradients of the fluid four-velocity. Now, for small
colliding systems, there is no clear separation between microscopic
and macroscopic scales, or the gradients are large, and yet
hydrodynamics seems to be capable of describing these systems. The
above observations suggest the existence of a new theory of
hydrodynamics that is applicable even far from local equilibrium
\cite{Romatschke:2017vte}. They have triggered a lot of theoretical
activity in recent years \cite{Florkowski:2017olj}.


\subsubsection{Anisotropic parton-escape models}

As illustrated in Fig. \ref{strickland}, due to the nonspherical shape
of the fireball, the parton escape probability depends on the angle at
which it is trying to escape, giving rise to an anisotropic momentum
distribution of the detected particles, without the need of any
pressure gradient. This has been demonstrated for p(d)-nucleus
collisions, within the framework of transport models (e.g., AMPT)
which by definition deal with dilute or low-density systems and allow
only a few scatterings of partons unlike in the hydrodynamical picture
\cite{He:2015hfa}. Apart from the anisotropic flow, its mass ordering
was also reproduced in these calculations \cite{Li:2016flp}. This
mechanism is expected to play an important role when the event
multiplicities are somewhat lower than those corresponding to the
hydrodynamic flow (Fig. \ref{strickland}). 
Interestingly, an early work that considered an expanding mixture of
several species of massive relativistic particles, in the framework of
the Boltzmann equation, concluded that a single collision per particle
on average already leads to sizeable elliptic flow, with mass
ordering between the species \cite{Borghini:2010hy}.
More recently, a simple kinetic theory
estimate showed that for small-enough systems, even a single-hit
dynamics is able to generate significant elliptic flow
\cite{Kurkela:2018qeb}.


\subsubsection{Colour-Glass-Condensate (CGC) effective field theory}

\begin{figure}
\begin{center}
\resizebox{0.4\columnwidth}{!}{\includegraphics{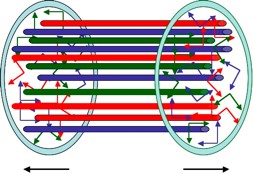}}
\caption{(colour online) Schematic representation of Glasma formed between two
  receding nuclei just after the collision. Figure from \cite{McLerran:2008es}.}
\label{glasma}
\end{center}
\end{figure}

This is an effective theory of QCD at high energy. Figure
\ref{strickland} shows where this approach is expected to play a
dominant role. Unlike hydrodynamics, this is not an
initial-spatial-geometry-driven approach. Models based on this
approach have intrinsic momentum-space correlations in the initial
multi-gluon distribution,
and hence, are called initial-state models. These correlations have a
similar structure as the experimentally observed correlations. The
essential physics idea is that the quarks from the small-ion
projectile scatter coherently off localized domains of strong
chromo-electromagnetic fields in the heavy-ion target. The colour
domains are of length scale $L \sim 1/Q_s$, where $Q_s$ is the
saturation scale in the target. This approach is able to reproduce
many of the features of multiparticle azimuthal correlations observed
in small systems
\cite{Dusling:2017dqg,Dusling:2017aot,Mace:2018vwq}. It has been
argued that the large $v_2$ for $J/\psi$ observed at LHC can be
naturally explained as the initial-state effect within the CGC
formalism \cite{Zhang:2019dth,Zhang:2020ayy}. Also, the glasma initial
conditions --- an ensemble of approximately boost-invariant flux tubes
of longitudinal colour electric and magnetic fields stretching between
the colour charges in the two receding nuclei (Fig. \ref{glasma}) ---
as predicted by CGC are consistent with the long-range structure seen
in the ridge phenomenon \cite{Dumitru:2010iy,Schenke:2016lrs}.

\bigskip

Thus the study of small systems not only deepens our understanding of
(hydro)dynamics at play in large systems, but also throws light on the
working of QCD in a many-body environment. Apart from
\cite{Strickland:2018exs}, here are some more recent reviews that
cover collectivity in small systems:
\cite{Ortiz:2019osu,Morrow:2018adw,Nagle:2018nvi,Schenke:2017bog,Li:2017qvf,Schlichting:2016sqo,Loizides:2016tew,Dusling:2015gta}.


\section{Conclusions}

As stated in the Introduction, this was not meant to be a
comprehensive review of the phenomenology of the collective flow
observed in relativistic collisions. Instead, the intention was to
provide the necessary mathematical background and explain key physics
concepts, in a pedagogical way, to someone uninitiated in the field,
so that they can follow the literature. To that end, interspersed
throughout the text are several exercises, which the reader is urged
to attempt.


\section{Acknowledgements}

I am very thankful to Jean-Yves Ollitrault for helpful comments on the
manuscript. I also thank him for our long-term collaboration which
allowed me to learn many things. I acknowledge the award of the Core
Research Grant, by the Science and Engineering Research Board,
Department of Science and Technology, Government of India. I thank
Bhavya Bhatt for drawing Fig. \ref{fig:skewkurt}.

\section{Appendix A}
\label{appA}

\setcounter{equation}{0}
\renewcommand{\theequation}{A\arabic{equation}}


\subsection{Moments and cumulants of a probability distribution}

The $n$-th moment of a (real, continuous) function $f(x)$, about a
constant $a$, is defined as
\begin{equation}
\mu_n(a) \equiv \int_{-\infty}^\infty (x-a)^n f(x) dx.
\end{equation}
We shall assume $f(x)$ to be the probability density function (PDF),
normalized to unity. The two most interesting values of $a$ are 0 and
$\mu\equiv \mean{x}$, the mean of the distribution. Usually one
refers to $\mu_n(a=0)$ simply as the ``moment'' and $\mu_n(a=\mu)$ as
the ``central moment''. Henceforth, we denote moments $\mu_n(a=0)$ by
$\mu'_n$ and central moments $\mu_n(a=\mu)$ by $\mu_n$. Obviously,
$\mu'_0=1=\mu_0$ and $\mu'_n=\mean{x^n}$. The first four central moments are
\begin{eqnarray}\left.\begin{aligned}
\mu_1&=0, \\
\mu_2&=\mean{x^2}-\mu^2 \equiv {\rm variance} \,(\sigma^2) \equiv \,\,
(\rm standard ~deviation = \sigma)^2, \\
\mu_3&= \mean{x^3}-3\mu \mean{x^2}+2\mu^3, \\
\mu_4&= \mean{x^4}-4\mu\mean{x^3}+6\mu^2 \mean{x^2}-3\mu^4.
\end{aligned}\right.\end{eqnarray}
It is often convenient to define standardized central moments which are
scale-invariant or dimensionless quantities:
$\mu_n/\sigma^n$. The first four standardized central moments are
\begin{eqnarray}\label{A3}
\left.\begin{aligned}
\mu_1/\sigma &= 0,  \\
\mu_2/\sigma^2 &= 1, \\
\mu_3/\sigma^3 &\equiv {\rm skewness} \,\, (\gamma), \\
\mu_4/\sigma^4 &\equiv {\rm kurtosis} \,\, (\kappa).
\end{aligned}\right.\end{eqnarray}
Excess kurtosis is defined as $\kappa-3$. However, it is not uncommon
to find the excess kurtosis itself termed as kurtosis. The reason for
subtracting 3 will become clear when we discuss the Gaussian (or
normal) distribution (Appendix C).

$\bullet$ Variance is a measure of the spread of the random numbers
about their mean value.

\begin{figure}
\centering
\includegraphics[width=0.38\textwidth]
{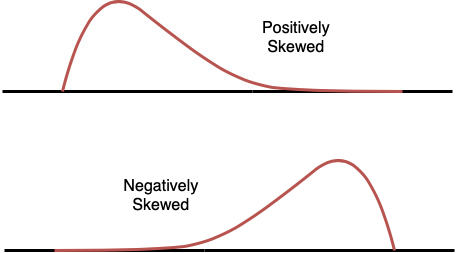}
\includegraphics[width=0.38\textwidth]
{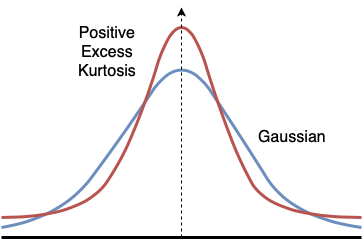}
\caption{(colour online) (left) Positively and negatively skewed distributions.
(right) Gaussian (excess kurtosis =
  0, blue) and a non-Gaussian distribution with heavier tails
  (excess kurtosis positive, red).}
\label{fig:skewkurt}
\end{figure}

$\bullet$ Skewness is a measure of the lopsidedness or asymmetry of
the distribution about its mean. It is clear from the definition that
a distribution that is symmetric about its mean has vanishing
skewness. In general, skewness can be positive or negative. If the
left (right) tail is drawn out, or in other words, is longer than the
right (left) tail, the distribution is said to be left(right)-skewed
and has a negative (positive) skewness; see Fig \ref{fig:skewkurt}.

$\bullet$ Kurtosis is a measure of the heaviness of the tails of the
distribution as compared to the normal distribution with the same
variance. It is clear from the definition that kurtosis is a
nonnegative number. (Excess kurtosis may be positive or negative.)
Kurtosis is a measure of the ``tailedness'' and
not the ``peakedness'' of a distribution (Fig \ref{fig:skewkurt}), 
because the proportion of
the kurtosis that is determined by the central $\mu \pm \sigma$ range
is usually quite small \cite{Westfall:2014phw}.

\subsection{Moment-generating function $M(t)$}

\begin{equation}
M(t)\equiv\mean{e^{tx}}=1+t\mean{x}+\frac{t^2}{2!}\mean{x^2}+\cdots = \sum_{n=0}^\infty \frac{t^n}{n!}\mean{x^n}.
\end{equation}
Observe that the moments $\mu'_n=\mean{x^n}$ appear in the above
expansion. To isolate the $m$-th moment $\mean{x^m}$, for example, one
uses
\begin{equation}
\left[\frac{d^m}{dt^m} M(t) \right]_{t=0} =\mean{x^m}.
\end{equation}
Differentiating $M(t)$ \, $m$ times removes the first $m$ terms, i.e.,
the terms containing $1, \mean{x}, \cdots, \mean{x^{m-1}}$. Further,
setting $t=0$ removes terms containing $\mean{x^{m+1}},\mean{x^{m+2}},$
$\cdots$, leaving behind only $\mean{x^m}$. The moment-generating
function provides an alternative description of the PDF.
The {\bf central moment generating function} is $e^{-\mu t}M(t)$.


\subsection{Cumulant-generating function $K(t)$}

\begin{equation}
K(t)\equiv \ln M(t) = \ln \mean{e^{tx}}
=t\kappa_1+\frac{t^2}{2!}\kappa_2+\frac{t^3}{3!}\kappa_3+\cdots = \sum_{n=1}^\infty \frac{t^n}{n!}\kappa_n,
\end{equation}
where $\kappa_n$ are the cumulants. To isolate the $m$-th cumulant
$\kappa_m$, for example, one uses
\begin{equation}
\left[\frac{d^m}{dt^m} K(t) \right]_{t=0} =\kappa_m.
\end{equation}
Differentiating $K(t)$ \, $m$ times removes the first $m-1$ terms, i.e.,
the terms containing $\kappa_1, \cdots, \kappa_{m-1}$. Further,
setting $t=0$ removes terms containing $\kappa_{m+1},\kappa_{m+2},
\cdots$, leaving behind only $\kappa_m$. It is straightforward to show
that the first few cumulants are
\begin{eqnarray}\left.\begin{aligned}
\kappa_1&= \mean{x} = {\rm mean} \,\, \mu, \\
\kappa_2&= \mean{x^2}-\mean{x}^2 = \mean{(x-\mean{x})^2}  
={\rm variance} \,\, (\sigma^2) = 
{\rm second \,\,central\,\, moment}, \\
\kappa_3&=
\mean{x^3}-3\mean{x}\mean{x^2}+2\mean{x}^3 = \mean{(x-\mean{x})^3} =
{\rm third \,\, central \,\, moment}, \\
\kappa_4&= \mean{x^4}-4\mean{x}\mean{x^3}-3\mean{x^2}^2+12\mean{x}^2\mean{x^2}
-6\mean{x}^4  \\
&\ne {\rm fourth \,\, central \,\, moment}
\,\,\mean{(x-\mean{x})^4}, \\
\kappa_5&= \mean{x^5} -5\mean{x}\mean{x^4} -10\mean{x^2}\mean{x^3}
+20\mean{x}^2\mean{x^3} +30\mean{x}\mean{x^2}^2 \\
& -60\mean{x}^3\mean{x^2} +24\mean{x}^5, \\
\kappa_6&= \mean{x^6}-6\mean{x}\mean{x^5}-15\mean{x^2}\mean{x^4}
+30\mean{x}^2\mean{x^4}-10\mean{x^3}^2
+120\mean{x}\mean{x^2}\mean{x^3} \\
&-120\mean{x}^3\mean{x^3}
+30\mean{x^2}^3
- 270\mean{x}^2\mean{x^2}^2
+360\mean{x}^4\mean{x^2}-120\mean{x}^6.
\end{aligned}\right.\end{eqnarray}
Fourth- and higher-order cumulants are not identical to the
corresponding central moments. Thus cumulants are certain polynomial
functions of the moments and provide an alternative to the moments of
the distribution.

The above equations can be inverted to express moments in terms of cumulants:
\begin{eqnarray}\left.\begin{aligned}
\mean{x} &= \kappa_1,  \\
\mean{x^2}&= \kappa_1^2+\kappa_2,  \\
\mean{x^3} &= \kappa_1^3+3\kappa_1\kappa_2+\kappa_3,  \\
\mean{x^4} &= \kappa_1^4+6\kappa_1^2\kappa_2+4\kappa_1\kappa_3+3\kappa_2^2+\kappa_4,
\\
\mean{x^5} &= \kappa_1^5+10\kappa_1^3\kappa_2+10\kappa_1^2\kappa_3
+15\kappa_1\kappa_2^2+5\kappa_1\kappa_4+10\kappa_2\kappa_3+\kappa_5,
\\
\mean{x^6} &= \kappa_1^6+6\kappa_1\kappa_5+15\kappa_1^2\kappa_4
+20\kappa_1^3\kappa_3+15\kappa_1^4\kappa_2+60\kappa_1\kappa_2\kappa_3
+45\kappa_1^2\kappa_2^2+15\kappa_2\kappa_4 \\
&+10\kappa_3^2+
15\kappa_2^3+\kappa_6.
\end{aligned}\right.\end{eqnarray}


\section{Appendix B}
\label{appB}

\setcounter{equation}{0}
\renewcommand{\theequation}{B\arabic{equation}}

\subsection{Correlation functions and cumulants}

The $n$-particle correlation function (or simply a correlator)
$\rho(1,2,3,\cdots,n)$ consists of terms that represent combinations
of lower-order correlations and a term that represents a genuine or
``true'' $n$-particle correlation $C(1,2,3,\cdots,n)$ which is called
a cumulant. For example, $\rho(1,2)=\rho(1)\rho(2)+C(1,2)$, so that
$C(1,2)=\rho(1,2)-\rho(1)\rho(2)$. If the two particles are
statistically independent, $\rho(1,2)$ simply reduces to
$\rho(1)\rho(2)$, whereas $C(1,2)$ vanishes by
construction.\footnote{For simplicity of notation, we use the same
  symbols $\rho$ and $C$ to denote 1-,2-,3- and multiparticle correlation
  functions and cumulants, respectively.} The first few correlation functions are
\cite{Bzdak:2015dja}
\begin{eqnarray}\left.\begin{aligned}
\rho(1)&=C(1), \\
\rho(1,2)&=\rho(1)\rho(2)+C(1,2), \\
\rho(1,2,3)&=\rho(1)\rho(2)\rho(3)+\rho(1)C(2,3)+\rho(2)C(3,1)+
\rho(3)C(1,2)+C(1,2,3),\\
&\equiv \rho(1)\rho(2)\rho(3)+\sum_{(3)}\rho(1)C(2,3)+C(1,2,3), \\ 
\rho(1,2,3,4)&=\rho(1)\rho(2)\rho(3)\rho(4)+\sum_{(6)}\rho(1)\rho(2)C(3,4)
+\sum_{(4)}\rho(1)C(2,3,4) \\
&+\sum_{(3)}C(1,2)C(3,4)+C(1,2,3,4), \\
\rho(1,2,3,4,5)&=\rho(1)\rho(2)\rho(3)\rho(4)\rho(5)
+\sum_{(10)}\rho(1)\rho(2)\rho(3)C(4,5) \\
&+\sum_{(10)}\rho(1)\rho(2)C(3,4,5)+\sum_{(15)}\rho(1)C(2,3)C(4,5) 
 \\
&+\sum_{(5)}\rho(1)C(2,3,4,5)
+\sum_{(10)}C(1,2)C(3,4,5)
+C(1,2,3,4,5), \\
\rho(1,2,3,4,5,6)&=\rho(1)\rho(2)\rho(3)\rho(4)\rho(5)\rho(6)
+\sum_{(6)}\rho(1)C(2,3,4,5,6) \\
&+\sum_{(15)}\rho(1)\rho(2)C(3,4,5,6)
+\sum_{(20)}\rho(1)\rho(2)\rho(3)C(4,5,6) \\
&+\sum_{(15)}\rho(1)\rho(2)\rho(3)\rho(4)C(5,6)
+\sum_{(60)}\rho(1)C(2,3)C(4,5,6) \\
&+\sum_{(45)}\rho(1)\rho(2)C(3,4)C(5,6)
+\sum_{(15)}C(1,2)C(3,4,5,6) \\
&+\sum_{(10)}C(1,2,3)C(4,5,6)
+\sum_{(15)}C(1,2)C(3,4)C(5,6) \\
&+C(1,2,3,4,5,6),
\end{aligned}\right.\end{eqnarray}
where the numbers in parentheses under the summation signs indicate the
number of possible permutations of the indices.

The above equations can be inverted to get the expressions for the cumulants
in terms of the correlation functions:
\begin{eqnarray}\left.\begin{aligned}
C(1)&=\rho(1), \\
C(1,2)&=\rho(1,2)-\rho(1)\rho(2), \\
C(1,2,3)&=\rho(1,2,3)-\sum_{(3)}\rho(1)\rho(2,3)+2\rho(1)\rho(2)\rho(3),
\\ 
C(1,2,3,4)&=\rho(1,2,3,4)-\sum_{(4)}\rho(1)\rho(2,3,4)
-\sum_{(3)}\rho(1,2)\rho(3,4) \\
&+2\sum_{(6)}\rho(1)\rho(2)\rho(3,4)
-6\rho(1)\rho(2)\rho(3)\rho(4), \\
C(1,2,3,4,5)&= \rho(1,2,3,4,5)
-\sum_{(5)}\rho(1)\rho(2,3,4,5)
-\sum_{(10)}\rho(1,2)\rho(3,4,5)\\
&+2\sum_{(10)}\rho(1)\rho(2)\rho(3,4,5)
+2\sum_{(15)}\rho(1)\rho(2,3)\rho(4,5) \\
&-6\sum_{(10)}\rho(1)\rho(2)\rho(3)\rho(4,5)
+24\rho(1)\rho(2)\rho(3)\rho(4)\rho(5), \\
C(1,2,3,4,5,6) &= \rho(1,2,3,4,5,6)
-\sum_{(6)}\rho(1)\rho(2,3,4,5,6)
-\sum_{(15)}\rho(1,2)\rho(3,4,5,6)  \\
&+2\sum_{(15)}\rho(1)\rho(2)\rho(3,4,5,6)
-\sum_{(10)}\rho(1,2,3)\rho(4,5,6) \\
&+4\sum_{(30)}\rho(1)\rho(2,3)\rho(4,5,6)
-6\sum_{(20)}\rho(1)\rho(2)\rho(3)\rho(4,5,6) \\
&+2\sum_{(15)}\rho(1,2)\rho(3,4)\rho(5,6)
-3\sum_{(90)}\rho(1)\rho(2)\rho(3,4)\rho(5,6) \\
&+24\sum_{(15)}\rho(1)\rho(2)\rho(3)\rho(4)\rho(5,6)
-\sum_{(120)}\rho(1)\rho(2)\rho(3)\rho(4)\rho(5)\rho(6).
\end{aligned}\right.\end{eqnarray}
It is easy to verify that the cumulant vanishes if any one or more
particles is statistically independent of the others. Thus the
$n$-particle cumulant measures the statistical dependence of the
entire $n$-particle set. For this reason, cumulants are also called
connected correlation functions. 

\bigskip

\noindent\fbox{%
    \parbox{\textwidth}{%
{\bf Exercise 11}: Compare the above expressions of
cumulants to those in Appendix A.
    }%
}

\bigskip

\noindent\fbox{%
    \parbox{\textwidth}{%
{\bf Exercise 12}: Note how the above expressions of correlators and
cumulants simplify considerably if the single-particle $\rho$
vanishes.
    }%
}


\section{Appendix C}
\label{appC}
\subsection{Gaussian or normal distribution in 1D}

\setcounter{equation}{0}
\renewcommand{\theequation}{C\arabic{equation}}

One-dimensional Gaussian (centered at $\mu$) is
\begin{equation}
f(x)=\frac{1}{\sigma \sqrt{2\pi}}\exp \left[-\frac{(x-\mu)^2}{2\sigma^2}\right],
\end{equation}
where $\mu=\mean{x}$ is the mean, $\sigma^2$ is the variance and
$f(x)$ is normalized to unity. The odd central moments are obviously
zero. The even central moments are given by $(n-1)!! \,\sigma^n \,\,
(n ~{\rm even})$. The first few even central moments are given in
Table 1. The skewness ($\gamma$) is zero, kurtosis ($\kappa$) is 3 and
excess kurtosis ($\kappa-3$) is zero. A probability distribution with
tails heavier (lighter) than those of the normal distribution shows a
higher (lower) propensity to produce outliers and has a positive
(negative) excess kurtosis  (Fig \ref{fig:skewkurt}),  

\begin{table}[h!]
  \begin{center}
    \caption{First few central and noncentral moments of the Gaussian distribution}
    \label{table1}
    \begin{tabular}{|c|l|l|}
\hline
Order & Central moment & Noncentral moment \\
\hline
1 & 0 & $\mu$ \\
2 & $\sigma^2$ & $\mu^2+\sigma^2$ \\
3 & 0 & $\mu^3+3 \mu \sigma^2$ \\
4 &  $3\, \sigma^4$ & $\mu^4+6 \mu^2 \sigma^2+3 \sigma^4$\\
5 & 0 & $\mu^5+10 \mu^3 \sigma^2+15 \mu\sigma^4$\\
6 & $15\, \sigma^6$ & $\mu^6+15 \mu^4 \sigma^2+45 \mu^2\sigma^4+15\sigma^6$\\
7 & 0 & $\mu^7+21 \mu^5 \sigma^2+105 \mu^3\sigma^4+105\mu\sigma^6 $\\
8 & $105\, \sigma^8$ & $\mu^8+28 \mu^6 \sigma^2+210 \mu^4\sigma^4+420\mu^2\sigma^6
+105\sigma^8 $\\
\hline
    \end{tabular}
  \end{center}
\end{table}

For the normal distribution, the moment-generating function is $M(t)=
\exp\,(\mu t + (\sigma^2 t^2 /2))$ and the cumulant-generating
function is $K(t) = \ln M(t)= \mu t + (\sigma^2 t^2 /2)$. Since this
is a quadratic in $t$, only the first two cumulants survive, namely
the mean $\mu$ and the variance $\sigma^2$.  It can be shown that the
normal distribution is the only one for which the third and higher
cumulants vanish.

\bigskip

\noindent\fbox{%
    \parbox{\textwidth}{%
{\bf Exercise 13}: Using the expressions for moments in terms of cumulants given 
in Appendix A, show that the first six noncentral moments of the normal distribution 
are as given in Table 1.
    }%
}


\subsection{Gaussian or normal distribution in 2D}

Two-dimensional Gaussian centered at the origin and normalized to
unity is
\begin{equation}
f(x,y)=\frac{1}{2 \pi \sigma_x \sigma_y} \exp \left[
-\frac{x^2}{2\sigma_x^2}-\frac{y^2}{2\sigma_y^2}
\right],
\end{equation}
where $\sigma_x^2$ and $\sigma_y^2$ are the variances. 

\bigskip

\noindent\fbox{%
    \parbox{\textwidth}{
{\bf Exercise 14}: For an asymmetric ($\sigma_x \ne \sigma_y$) 2D
Gaussian as in Eq. (C2), let $z=x+iy$. Show that the fourth moment $\langle
z^4\rangle \ne 0$, and is trivially correlated with the second moment
$\langle z^2\rangle$. However, the fourth-order cumulant $\langle
z^4\rangle - 3 \langle z^2\rangle^2=0$.
}
}

\bigskip

It is often
convenient to introduce the symmetric case ($\sigma_x=\sigma_y$) and
write it in terms of $r=\sqrt{x^2+y^2}$:
\begin{equation}
f(r)
=\frac{1}{\pi \sigma^2}\exp \left[-\frac{r^2}{\sigma^2}\right],
\end{equation}
where $\sigma^2\equiv\sigma_x^2+\sigma_y^2$. $f(r)$ is also centered
at the origin and normalized to unity. Note, however, that unlike
$\mean{x}$ and $\mean{y}$, $\mean{r}$ is not zero. The first few
moments of $f(r)$, $\mu_n=\mean{r^n}=\sigma^n \, \Gamma((n/2)+1)$, are
given in Table 2. Note also that $\mean{r^2}=\mean{x^2}+\mean{y^2}$;
using $\mean{r^2}$ from Table 2, we get
$\sigma^2=\sigma_x^2+\sigma_y^2$. The kurtosis in this case is 2,
unlike the case of a 1D Gaussian discussed earlier where it was 3.
\begin{table}[h!]
  \begin{center}
    \caption{First few moments ($\mu_n$) of $f(r)$, Eq. (C3)}
    \label{table2}
    \begin{tabular}{|c|c|c|c|c|c|c|c|c|}
\hline
      $n$ & 1 & 2 & 3 & 4 & 5 & 6 & 7 & 8 \\
\hline
$\mu_n$ & $\sqrt{\pi}\,\sigma/2$ & $\sigma^2$ &
$3\sqrt{\pi}\,\sigma^3/4$ & $2\, \sigma^4$ & 
$15 \sqrt{\pi}\,\sigma^5/8$ & $6\, \sigma^6$ & 
$105 \sqrt{\pi}\,\sigma^7/16$ &
$24\, \sigma^8$ \\
\hline
    \end{tabular}
  \end{center}
\end{table}


\end{document}